\documentclass[12pt,a4paper]{article}

\usepackage{a4}
\usepackage{amsfonts}
\usepackage{amssymb}
\usepackage{epsfig}
\usepackage{psfig}
\usepackage{psfrag}
\usepackage{dsfont}

\newcommand{\beqn}{\begin{eqnarray}}
\newcommand{\eeqn}{\end{eqnarray}}
\newcommand{\be}{\begin{equation}}
\newcommand{\ee}{\end{equation}}
\newcommand{\non}{\nonumber \\}

\begin{document}

\title{}
\begin{flushright}
\vspace{-3cm}
{\small MIT-CTP-3418 \\
        NUB-TH- 3240\\
        hep-th/0309167
 }
\end{flushright}
\vspace{1cm}

\begin{center}
{\Large\bf 
Effective Action and \\[.1cm] Soft Supersymmetry Breaking \\[.3cm] for Intersecting D-brane Models}
\end{center}

\vspace{1.5cm}

\author{}
\date{}
\thispagestyle{empty}

\begin{center}

{\bf Boris K\"ors}\footnote{e-mail: kors@lns.mit.edu}$^{,*}$
{\bf and Pran Nath}\footnote{e-mail: nath@neu.edu}$^{,\dag}$
\vspace{.5cm}

\hbox{
\parbox{8cm}{
\begin{center}
{\it
$^*$Center for Theoretical Physics \\ 
Laboratory for Nuclear Science \\ 
and Department of Physics \\ 
Massachusetts Institute of Technology \\ 
Cambridge, Massachusetts 02139, USA \\
}
\end{center}
} 
\hspace{-.5cm}
\parbox{8cm}{\begin{center}
{\it
$^\dag$Department of Physics \\ 
Northeastern University \\
Boston, Massachusetts 02115, \\ USA \\
}
\end{center}
}
}

\vspace{.5cm}
\end{center}

\begin{center}
{\bf Abstract} \\
\end{center}

We consider a generic scenario of spontaneous breaking of supersymmetry in the hidden sector within 
${\cal N}=1$ supersymmetric orientifold compactifications of type II string theories with 
D-branes that support semi-realistic chiral gauge theories. 
The soft breaking terms in the visible sector of the models are computed in a standard way without specifying 
the breaking mechanism, which leads to expressions that generalize those formerly known for heterotic or 
type I string models. 
The elements of the effective tree level supergravity action relevant for this, such as the 
K\"ahler metric for the matter fields, the superpotential of the visible sector and the gauge kinetic functions, 
are specified by dimensional reduction and duality arguments. As phenomenological applications 
we argue that gauge coupling unification can only occur in special regions of the
moduli space; we show that flavor changing neutral currents can be suppressed sufficiently for a wide range of parameters, 
and  we briefly address the issues of CP violation, electric dipole moments and dark matter, as well. 

\clearpage
\setcounter{footnote}{0}

\section{Introduction} 

String theory has produced a variety of approaches to construct models 
that come rather close to the 
qualitative features of low energy particle physics. While the first such models
 were based on the heterotic 
string \cite{heterotic}, the advent of D-branes has allowed for a broader perspective and more direct bottom-up attempts 
to meet the desired gauge group, chiral matter spectrum etc. An approach that is maybe distinguished by 
its computability and simplicity together with very appealing phenomenological 
possibilities is that of 
type IIA orientifold compactifications (see \cite{Angelantonj:2002ct} for a general review)  
on Calabi-Yau (CY) manifolds with intersecting D6-branes. 
It can produce a plethora of 
models with finite effective four-dimensional Planck-scale, with ${\cal N}=1$ supersymmetry in 
four dimensions, gauge groups that 
include or consist of products of unitary groups, and chiral fermionic matter in the form of bifundamental 
representations - all the ingredients the supersymmetric extensions of the Standard Model 
(we shall often just refer to the MSSM) need \cite{Blumenhagen:1999md,Blumenhagen:1999ev,
Pradisi:1999ii,Blumenhagen:1999db,Angelantonj:2000hi,
Forste:2000hx,Angelantonj:2000rw,Forste:2001gb,Blumenhagen:2001te,Cvetic:2001tj,Cvetic:2001nr,
Blumenhagen:2002wn,Blumenhagen:2002gw,Honecker:2003vq,Li:2003xb,Larosa:2003mz}. The first semi-realistic 
supersymmetric models that come close to the MSSM spectrum and gauge group were presented 
in \cite{Cvetic:2001tj,Cvetic:2001nr}, which we will frequently 
refer to as prominent examples.\footnote{An extensive review of the 
CY-orbifold models with intersecting D-branes, that contains a 
more complete introduction and nice illustrations has recently appeared in \cite{Ott:2003yv}.}
A point of contact between string models and testable physics comes from
the soft breaking terms and we will follow here the hidden sector 
scenarios of SUGRA models in deducing these \cite{Chamseddine:jx,Barbieri:1982eh,Hall:iz, Nath:aw}. \\ 

In a first approach the attention in intersecting brane models 
has been concentrated on models which break supersymmtry already at the string 
scale \cite{Blumenhagen:2000wh,Blumenhagen:2000vk,Aldazabal:2000cn,Aldazabal:2000dg,
Blumenhagen:2000ea,Ibanez:2001nd,Kokorelis:2002ip,Kokorelis:2002zz,Kokorelis:2002wa}, 
and may thus be considered phenomenologically relevant only in the framework of low string scale 
or large extra dimension scenarios \cite{Arkani-Hamed:1998rs,Antoniadis:1998ig}. 
This is mainly due to the fact that supersymmetry imposes extra 
conditions, as it turnes out conditions that correspond to the vanishing of Fayet-Iliopolous-terms 
in the effective theory \cite{Berkooz:1996km,Cremades:2002te,Blumenhagen:2002wn}. 
These make it much harder to construct interesting models in an economical fashion, 
but some progress has been made, see e.g. \cite{Cvetic:2002qa,Cvetic:2002wh,Cvetic:2002pj,
Cremades:2003qj,Cvetic:2003xs,Blumenhagen:2003jy}. Of course, this also leaves the question open, how supersymmetry gets finally 
broken in these models. Circumventing this question for the moment, we examine the 
consequences of a rather generic scenario where a hidden sector breaks supersymmtry in some unspecified fashion. 
To do so, we shall have to determine various elements of the effective four-dimensional low energy Lagrangian, such 
as parts of the K\"ahler potential, the superpotential, gauge kinetic functions etc. We do this partly by direct 
dimensional reduction and partly by invoking the perturbative duality between the type I and heterotic 
string in four dimensions. Ultimately, the goal of this program is to put the phenomenological models 
based on type I strings and D-branes on the same footing with heterotic models studied in the past. \\ 

Concretely, we assume the model to contain two 
sectors, a visible sector that contains at least the MSSM matter fields, and one or more hidden sectors left unspecified. 
Supersymmetry is then assumed to be broken in a fashion that is standard in supergravity models 
(see \cite{Nath:1983fp,Nilles:1983ge,Kaplunovsky:1993rd} and references therein):  
through some unknown, possibly non-perturbative, mechanism in the hidden sector a potential is generated 
at some intermediate scale, usually taken to be around $M_{\rm sb} \sim 
10^{13}$GeV, and supersymmetry breaks down spontaneously. 
This induces explicit but soft breaking in the visible sector. 
As an example, in 
\cite{Cvetic:2003yd} the possibility of gaugino condensation in the hidden sector was considered,\footnote{Actually, 
\cite{Cvetic:2003yd} 
relied on slightly different assumptions about the breaking patterns and the moduli sector, which e.g. lead to a large 
negative cosmological constant. This does not necessarily appear in our more complete treatment.} but in principle one may also 
want to allow mechanisms that are not based on the dynamics of gauge fields and thus do not involve further hidden D-branes.   
An alternative example could be background fluxes that generate a perturbative scalar potential. 
We shall in fact leave the concrete breaking mechanism open, much in the spirit of \cite{Kaplunovsky:1993rd}, 
and only parametrize the breaking. We then compute the soft breaking 
parameters and discuss various issues that arise. There are many phenomenological restrictions that apply to 
the soft breaking terms in the effective action, and just to show this in some examples, we discuss the issues of 
flavor changing neutral currents (FCNC), CP violation, electric dipole moments (EDM), and dark matter. 
We also look at the perspectives to achieve a unification of 
gauge couplings in these models (see e.g. \cite{Blumenhagen:2003jy,Blumenhagen:2003qd,Chamoun:2003pf}),  
in order to derive  constraints on the 
moduli space for the gauge coupling unification to occur. We find that it requires rather special 
conditions on the moduli to be fulfilled and is not generic. \\ 

We begin with a  technical remark on the practical formulation of the models: The more intuitive type IIA 
version of the models describes them as orientifolds of type IIA with D6-branes and 
orientifold 6-planes (O6-planes), filling out 3+1 dimensional Minkowski space-time and 
wrapping internal 3-manifolds. The intersection of these branes contain chiral fermionsin bifundamental representations
allowing one to attempt Standard Model like constructions. These models are thus apporiately called  the
intersecting brane models.
However, for our purposes it will be very helpful to rephrase the construction in terms of 
a type IIB orientifold with magnetic background fields on D9-branes (as in most of the original early 
works as \cite{Blumenhagen:2000fp,Angelantonj:2000hi,Blumenhagen:2000ea}). The physics of the two pictures is identical, 
since both are related by a simple T-duality along three directions of the internal space, actually mirror symmetry. 
The advantage of the latter 
formulation lies in the fact that it can directly be understood as a compactification of type I string theory 
only including background fluxes for the Yang-Mills field strengths on the D9-brane world volumes. This allows to 
determine various quantities either by dimensional reduction from known ten-dimensional type I expressions 
or from heterotic-type I duality. In the main text we shall refer to the two pictures as type IIA and type IIB versions 
and frequently employ both points of view interchangeably, since they are ultimately equivalent and hope that this will
not confuse the reader. \\ 

The complete picture that we put forward consists of the following elements: We compactify type I strings on a 
CY-orbifold space (type IIB version), add magnetic world volume fluxes, which introduces a new mass scale, 
and finally break supersymmetry spontaneously at the breaking scale $M_{\rm sb}$. 
The effective action which is obtained thereafter in principle implies that we have integrated out the massive string excitations, 
the massive Kaluza-Klein (KK) modes, those fields that get massive upon introducing the magnetic flux, and finally the 
hidden sector fields that decouple when supersymmetry is broken. The four relevant mass scales are given by the 
string scale $M_s$, the KK scale $1/R$, with $R$ an average radius, the magnetic flux scale $1/(M_s R^2)$ \cite{Kaloper:1999yr}, 
and the breaking scale 
$M_{\rm sb}$. For a space that is not ``isotropic'' it could also become necessary to include various different 
KK scales, such as for large transverse volume models.  
Since we are working in the supergravity approximation of large $R$, the splitting 
\be \label{scales} 
M_s \gg \frac{1}{R} \gg \frac{1}{M_sR^2} \gg 1{\rm TeV} 
\ee
is automatically implied for the self-consistency of the expansion of the effective action in derivatives. On the 
other hand, supersymmetry could be broken either below or above $1/(M_sR^2)$. The relevance of Eq.(\ref{scales}) for writing 
an effective action lies in the fact that it allows to treat the effects of the world volume fluxes as rather 
small perturbations of the background geometry, the solution to the vacuum equations of motion without fluxes. 
It would of course be very interesting, if one could go beyond this approximation in some controllable manner, 
but we shall stick to the ``probe limit'' in the following, neglecting any backreaction on the geometry. \\ 

   Before going into a detailed disucussion of the analysis, we describe 
   below 
   the main results of this paper. In previous works on intersecting brane
   models the Kahler potential of models  has not been fully specified.
   In this paper this potential is obtained and Eqs.(47) and (57) constitute
   two of the important results of this paper. This potential
   is then utilized to compute soft breaking for a generic class of D brane
   models. The soft breaking formula obtained in the paper are general 
   and encompass a large  class of D brane models and thus have a wide range 
   of applicability. Remarkably the full information 
   on soft breaking in these formulae is encoded in only a few
   indices which are determined in terms of the wrapping numbers.
   As a check on our results it is shown that the formula obtained 
   reproduce the previouslly known results for soft breaking in the  
   parallel brane case under specific limiting conditions.
   Further, an application of the general formulae for a specific intersecting
   D brane model involving D9 branes and D5 brane which contains
   perturbative, nonperturbative and interpolating  sectors is
   also made. Thus the soft breaking formulae given
   in Sec.4 encode information on soft breaking on a variety
   of D brane models. Further, a variety of other phenomena associated with
  intersecting D brane models are also discussed. Thus a significant 
  question in D brane models centers around unification of gauge 
  coupling constants. This issue has been addressed in the context of
   specific models in several recent papers. In this paper we show
   that the unification of gauge couplings is intimately tied to the
   constraints on the moduli needed to achieve ${\cal N}=1$ supersymmetry
   in intersecting D brane models. The new results of the paper are 
   contained in Secs. 3, 4 and 5 and in Appendix A. \\ 

   Next we compare and constrast the main results achieved in this paper
   with those of the heterotic strings.
Concretely, we will derive formulas for the K\"ahler metric of the open string fields with ends 
on the various stacks of D-branes, labelled by letters $a,b$, which turn out as 
\beqn  
\tilde{\cal K}_{m\bar m}^{[aa]}
&=&
\frac{1}{(s+ \bar s)(t_m + \bar t_{\bar m})(u_m + \bar u_{\bar m})} \frac{4\Re (f_a)}{1 + \Delta_a^{(m)}} \ , 
\non
\tilde{\cal K}^{[ab]}_{\alpha\bar\beta} &=&
\delta_{\alpha\bar\beta} (s+\bar s)^{\nu_{ab}/2-1}
 \prod_{m=1}^3 ( t_m +\bar t_{\bar m} )^{\nu^{(m)}_{ab}-\nu_{ab}/2}
 \prod_{m=1}^3 ( u_m +\bar u_{\bar m} )^{\nu^{(m)}_{ab}-\nu_{ab}/2}  \ . 
\nonumber
\eeqn
Here $\{s,t_m,u_m\}$ are the moduli fields, $f_a$ gauge kinetic functions, $\Delta_a^{(m)}$ some 
function of the moduli, and $\nu_{ab}$ numerical parameters. These two expressions are cearly distinguished from any 
K\"ahler metric known formerly for heterotic or type I compactifications. The first line refers to the $m$-component of 
open strings with both ends on the brane stack $a$, and it generalizes the known metric for untwisted heterotic fields, 
which reads 
\beqn 
\tilde{\cal K}_{m\bar m}^{\rm het}
&=&
\frac{1}{(t_m + \bar t_{\bar m})(u_m + \bar u_{\bar m})} \ , 
\nonumber
\eeqn 
while the limit $4\Re(f_a)/(1 + \Delta_a^{(m)}) = (s+\bar s)$ in which it reduces to this expression 
corresponds to a non-chiral limit of the D-brane models. The second line stands for open strings that connect branes $a$ and $b$. It 
has strong resemblence to twisted heterotic 
fields, but is as well clearly identified by the appearance of the dilaton $s+\bar s$ in the K\"ahler metric, which is 
impossible for the heterotic string. So for instance, a scenario of total dilaton dominance in the soft breaking, which leads to 
great simplification for heterotic models, is by far not as simple in the present class of D-brane compactifications. 
To make the above expressions practically useful, we further compute the soft breaking terms, where the novel K\"ahler metric leads 
to some new effects, such as for example ``interference effects'' of various moduli fields in the squark masses. In any case, 
our formulas allow a straightforward phenomenological interpretation for any brane model of the present type, just plugging 
in the parameters that characterize the particular model. The well-known phenomenological consequences of the soft 
breaking terms in the effective Lagrangian lead finally to restrictions on these parameters in order not to be 
in contradiction with current experimental bounds, which we exemplify by analyzing the appearance of falvour 
changing neutral currents. As a by product, we also give a systematic discussion of the perspectives to achieve a 
unification of gauge couplings, which turns out to be completely different approach than for the heterotic string once more, 
where it was rather automatic to achieve grand unification. \\ 

The organization of the paper is as follows: In section 2 we give an introduction to the relevant aspects of orientifolds 
with intersecting branes or branes with magnetic world volume fluxes, which we (not quite successfully) 
tried to keep short, and also introduce our conventions and notations for the effective Lagrangian. In section 3 we 
determine the necessary ingredients for using the effective action, parts of the K\"ahler potential, the superpotential, 
D-terms and FI parameters, as well as axionic couplings within St\"uckelberg mass terms. In section 4 we use these 
to compute the soft breaking terms in a rather straightforward manner, and discuss some implications. 
Finally, in section 5 we address the mentioned phenomenological issues, applying the expressions for the soft breaking terms. 


\section{Intersecting brane models on CY-orbifolds} 

We first like to specify the type of model we are considering and set up some notations and conventions. 
The reader who is familiar with the literature on intersecting brane models may even want to skip the section. 

\subsection{Definition of the class of models} 

Specifically, we are discussing toroidal CY-orbifold compactifications of type I string theory, 
or CY-orientifolds of type IIA, the latter version featuring D6-branes that intersect each other 
in points on an internal six-dimensional space \cite{Berkooz:1996km}. 
These models are known as intersecting brane world orbifolds. 
Their massless chiral fermion spectra can be engineered to match the Standard Model spectrum or that of 
grand unified theories. We shall actually 
concentrate on models that allow vacua that preserve exactly ${\cal N}=1$ supersymmetry in the effective 
four-dimensional action. The examples discussed explicitly in the literature have orbifold groups 
$\mathbb{Z}_2\times \mathbb{Z}_2$ \cite{Cvetic:2001nr}, $\mathbb{Z}_4$ \cite{Blumenhagen:2002gw}, or 
$\mathbb{Z}_4\times \mathbb{Z}_2$ \cite{Honecker:2003vq}, but all groups of 
even order are believed to allow supersymmetric vacua 
(based on the observation of \cite{Blumenhagen:2000ea}).\footnote{There has also been a simplified ``local'' approach 
to supersymmetric model building, that consists in relaxing the tadpole constraints, which we mention later, 
and just constructing a supersymmetric subsector of the whole model to produce the supersymmetric Standard Model, e.g. 
recently in \cite{Kokorelis:2003jr}.
The full model would however violate supersymmetry in these settings without the orbifold.} 
As mentioned in the introduction, 
we will not be directly using the language of intersecting brane models as type IIA orientifolds, but 
a T-dual description in the form of generalized orientifolds of type IIB string theory. The main modification 
compared to standard orientifolds \cite{Angelantonj:2002ct} then consists in 
allowing for non-trivial background fields on the world volume of the D-branes by 
including constant background values for gauge field strengths ${\cal F}$. Roughly speaking, T-duality relates 
the background gauge fields to the relative angles $\varphi$ among intersecting 
D6-branes in the type IIA intersecting brane model by a formula symbolically ${\cal F} = \tan (\varphi)$. \\ 

The full orientifold group $\{ \Omega{\cal R} \Theta^k, \Theta^l\}_{k,l\in \mathbb{Z}_N}$, 
that is divided out of the type IIA theory version of the scenario is given by the generator 
$\Theta$ of a the standard $\mathbb{Z}_N$ (or by two generators $\Theta_i$ in $\mathbb{Z}_N\times\mathbb{Z}_M,\ i=1,2$) 
toroidal orbifold group with crystallographic action on the background torus $\mathbb{T}^6$ \cite{Dixon:jw,Dixon:1986jc}, 
and by the modified world sheet parity $\Omega {\cal R}$. 
 We assume the background torus  to factorize into $\mathbb{T}^6 = (\mathbb{T}^2)^3$, so that the internal 
metric $G$ splits into 
\be 
G = {\rm diag} ( G^{(1)}, G^{(2)}, G^{(3)} )  \ , 
\ee
with each $G^{(m)}$ defining a $2\times 2$ metric on $\mathbb{T}^2_{m},\ m=1,2,3$. 
This ansatz excludes $\{ \mathbb{Z}_3, \mathbb{Z}_4,\mathbb{Z}_6'\}$ from our analysis, 
which do have off-diagonal moduli.   
We denote the co-ordinates in the three complex planes by $z_m = (i x_{2m-1} + U_m x_{2m})/\sqrt{2}$ 
and the standard complex moduli of any one of the three $\mathbb{T}^2_{m}$ by 
\be
T_m = {\rm vol} (\mathbb{T}^2_m) - i \int_{\mathbb{T}^2_m} B_2 = \sqrt{G^{(m)}} - i b^{(m)} \ ,\quad 
U_m = -i \frac{R^{(m)}_2}{R^{(m)}_1} e^{i \vartheta^{(m)}} = \frac{\sqrt{G^{(m)}} - iG^{(m)}_{12}}{G^{(m)}_{11}} \  . 
\ee
The slightly unconventional factor of $-i$ in the definition of the complex structure is 
introduced to make conventions compatible with \cite{LopesCardoso:1994is}.\footnote{The 
mapping of coordinates actually is $x=-ix',\ y=iy',\ U=-iU'$, the primed quantities referring to the standard torus 
conventions.} 
The $T_m$ and $U_m$ capture the two radii $R_i^{(m)}$ and the tilting angle $\vartheta^{(m)}$ 
as well as the component of the 
NSNS 2-form $B$-field $B^{(m)}_{12} = b^{(m)}$ along the respective torus. For the metric of the four-dimensional 
space-time we use $g_{\mu\nu}$, reserving indices $g_{ij}=G_{ij}$ for internal components. 
In terms of the moduli parameters the metric $G$ is expressed  
through the zwei-bein 
\beqn
{\bf e}_{1 a} = (R_1,0)_a\ ,\quad  
{\bf e}_{2 a} = (R_2 \cos(\vartheta), R_2 \sin(\vartheta))_a 
\eeqn
as $G^{(m)}_{ij} = \delta^{ab} {\bf e}_{i a} {\bf e}_{j b}$, or 
\beqn
G_{ij}^{(m)} = 
\frac{\Re(T_m)}{\Re(U_m)} 
\left( 
\begin{array}{c c}
1 & -\Im(U_m) \\ -\Im(U_m) & |U|^2 
\end{array}
\right)_{ij} 
\eeqn
With these conventions the operation ${\cal R}$ that appears in $\Omega{\cal R}$ 
can be taken the reflection of the real parts of the complex coordinates 
$z_m$ along $\mathbb{T}^2_m$, ${\cal R} : z_m \mapsto - \bar z_m$. The 
T-duality that translates this scenario into the more standard type I language, the type IIB version, 
is an inversion of the radii $R_1^{(m)}$ (actually the mirror symmetry transformation) 
and takes $\Omega {\cal R} \mapsto \Omega$ and $\Theta \mapsto \hat \Theta$, the latter being an asymmetric rotation that 
rotates left- and right-moving world sheet fields with opposite phases \cite{Blumenhagen:2000fp}. 
The duality also swaps the moduli, 
$U_m \mapsto T_m,\ T_m \mapsto U_m$. Therefore, whenever the requirement to have a crystallographic action of 
$\Theta_i$ has fixed the complex structure modulus $U_m$ of the type IIA background, 
$\hat\Theta_i$ now fixes the corresponding K\"ahler parameter $T_m$ in type IIB. 
Note that we will not use different notation for the type IIA and type IIB moduli, and our later notations will 
always refer to the type IIB version. The above implies that the dual model is an 
asymmetric orientifold which is mirror symmetric to the type IIA model and thus has swapped numbers of complex structure 
and K\"ahler moduli fields in its spectrum. For the untwisted fields, this means we are dealing with models that have the 
generic 3 complex structure moduli $U_m$ and 0, 1 or 3 K\"ahler moduli $T_m$, 
the latter case referring to $\mathbb{Z}_2\times \mathbb{Z}_2$ (see \cite{Ibanez:1992hc} for a list of orbifold moduli spaces). 
Note also that the imaginary 
parts of the $T_m$ are fixed through the orientifold projection $\Omega$ anyway such that $b^{(m)} = 0$ or $1/2$ 
\cite{Angelantonj:1999jh,Blumenhagen:2000ea}. 
The axionic imaginary parts of the scalar fields $s$ and $t_m$ 
in the chiral multiplets are in fact RR scalars that descend from reducing 
the RR potentials $C_2$ or $C_6$ instead of $B_2$ \cite{Cremades:2002te}. \\ 

It has been shown that supersymmetric ground states in the effective theory require the presence of orientifold 9-planes 
(O9-planes) together with O5-planes in order to be able to achieve complete tadpole cancellation. This is possible 
whenever the order of the orbifold group generators is even, and the most interest has by now been paid to 
the simplest example of the $\mathbb{Z}_2\times \mathbb{Z}_2$ model \cite{Cvetic:2001nr}. 
As mentioned above, this case is the most generic 
one in the sense that we will have to deal with completely generic $\mathbb{T}^2_m$ each with non-trivial complex 
structure and K\"ahler modulus. \\ 

In the type IIA picture the intersecting brane scenario was established by noticing that the charges and the tensions of the 
orientifold 6-planes, defined as the fixed locus of $\Omega{\cal R}$ on the orbifold, 
could be canceled by D6-branes which wrap on circles on any of the three $\mathbb{T}^2_m$ and 
fill out the four-dimensional Minkowski space-time \cite{Blumenhagen:2000wh}. On any such two torus a stack of $2N_a$ parallel 
D6$_a$-branes (a line) is defined by a set of two co-prime integers $({\bf n}_a^{(m)},\bar{\bf m}_a^{(m)})$, allowing 
$(1,0)$ and $(0,1)$ but no multiples thereof, denoting the lattice point 
which they meet, if drawn from the origin. 
The gauge group for the orbifold group $\mathbb{Z}_2\times \mathbb{Z}_2$ is given by 
\be \label{gaugegr} 
H = \bigotimes_a U(N_a) \ , 
\ee
with some $U(1)$ factors decoupling through axionic Green-Schwarz type couplings, and neglecting the option 
of orthogonal and symplectic gauge group factors. The primordial gauge group $U(2N_a)$ 
on $2N_a$ brane is broken by the orientifold projection, which can be demonstrated by regarding $N_a=1$ without 
loss of generality \cite{Douglas:1998xa}. 
First of all, the effect of $\Omega{\cal R}$ is to identify the stack of branes with 
winding numbers $({\bf n}_a^{(m)},\bar{\bf m}_a^{(m)})$ with the stack $({\bf n}_a^{(m)},-\bar{\bf m}_a^{(m)})$ and 
leaves the full $U(2)$. Choosing a basis of the adjoint by using $\sigma^0={\bf 1}_2$, and 
$\sigma^A,\ A=1,2,3,$ the Pauli-matrices, strings with both ends on either one of the branes can be identified with 
the component $(\sigma^0 \pm \sigma^3)/2$ while open strings between the two are in the directions $\sigma^{1,2}$. 
Now one generator of $\mathbb{Z}_2\times \mathbb{Z}_2$ projects out the components $\sigma^{1,2}$ of the gauge bosons 
$A_\mu^A \sigma^A$ and the other one identifies the two $(\sigma^0 \pm \sigma^3)/2$, 
such that one is left with a single $U(1)_a$ gauge factor.  
The most general solution is obtained by tensor products of these Chan-Paton matrices \cite{Douglas:1998xa}. 
For fields $\Phi^A\sigma^A$ that have opposite space-time parity under 
the generators of $\mathbb{Z}_2\times \mathbb{Z}_2$, i.e. $\Theta_i: \Phi^A \mapsto -\Phi^A$, 
exactly the opposite mapping applies and bifundamental fields 
of different gauge factors can survive in the spectrum. The same reasoning can also be applied to the 
$\mathbb{Z}_2 \times \mathbb{Z}_4$ orbifold, while the patterns for $\mathbb{Z}_4$ would be slightly different. 
As stressed in \cite{Blumenhagen:2002wn}, 
the projection on the Chan-Paton factors has a geometric interpretation in the blown-up 
version of the orbifolds. One should also mention that it is in fact possible to get $SO(N_a)$ and $Sp(N_a)$ 
gauge groups, which  have played a role in some applications \cite{Blumenhagen:2003jy,Blumenhagen:2003qd,Chamoun:2003pf}.  
For our purposes it will mostly suffice to treat the stacks 
of branes as supporting the factors of $H$, and the bifundamental part of the chiral matter spectrum given by their 
intersection numbers 
\be
I_{ab} = \prod_{m=1}^3 \left( {\bf n}_a^{(m)} {\bf m}_b^{(m)} - {\bf n}_b^{(m)} {\bf m}_a^{(m)} \right) \ , 
\ee
where ${\bf m}_a^{(m)} = \bar{\bf m}_a^{(m)} + b^{(m)} {\bf n}_a^{(m)}$. The realization of the Standard Model 
spectrum needs at least four different stacks of branes with primordial gauge group 
$U(3)\times U(2)\times U(1)\times U(1)$. The three generations of 
all the matter fields, quarks and leptons, can arise from multiple intersections of the same two stacks, respectively, 
except for the quark doublets, which split into two plus one from two different types of intersections in a minimal 
setting \cite{Ibanez:2001nd}. 
Beyond the bifundamental matter there are also chiral multiplets in symmetric and anti-symmetric representations, 
which are often completely ignored in a somewhat inconsistent manner. 
For a more complete treatment of this and the other models we refer to the original literature. \\ 

In the type IIB dual picture all the branes map into stacks of 
D9-branes (except for certain degenerate cases, which will be dealt with later) that fill out the entire ten-dimensional 
space-time but carry non-trivial background gauge flux on their respective world volume, given by 
\beqn 
{\cal F}_a &=& {\rm diag} ( {\cal F}^{(1)}_a , {\cal F}^{(2)}_a , {\cal F}^{(3)}_a ) \ , \non 
\left( {\cal F}^{(m)}_a \right)_{ij} &=& \left( b^{(m)} + \frac{\bar{\bf m}_a^{(m)}}{{\bf n}_a^{(m)}} \right) \epsilon_{ij} 
  = \frac{{\bf m}_a^{(m)}}{{\bf n}_a^{(m)}} \epsilon_{ij} = {\bf F}^{(m)}_a\,  \epsilon_{ij} \ ,\quad  i,j=1,2 \ . 
\eeqn
The gauge flux also factorizes into 2-tori, thus ${\cal F}_a$ is a $(1,1)$ form in complex notation, 
and the individual D9$_a$-branes carry labels $a$. 
We have made the field strength dimensionless by setting $2\pi \alpha' = 1$, which otherwise appears multiplying ${\cal F}$. 
Through the presence of $b^{(m)}$ the effective quantum number ${\bf m}_a^{(m)}$ may then be integer or half-integer.
The symmetry of the brane spectrum under $\Omega{\cal R}$ already implies the cancellation of half of all 
the RR charges, while the remaining conditions read 
\beqn \label{rrtad}
&& \sum_a N_a \prod_{m=1}^3 {\bf n}_a^{(m)} + N_{{\rm O9}} = 0 \ , \quad 
\sum_a N_a {\bf F}_a^{(1)} {\bf F}_a^{(2)} \prod_{m=1}^3 {\bf n}_a^{(m)} + N_{{\rm O5}_3} = 0 \ , \\
&& \sum_a N_a {\bf F}_a^{(1)} {\bf F}_a^{(3)} \prod_{m=1}^3 {\bf n}_a^{(m)} + N_{{\rm O5}_2} = 0 \ , \quad
\sum_a N_a {\bf F}_a^{(2)} {\bf F}_a^{(3)} \prod_{m=1}^3 {\bf n}_a^{(m)} + N_{{\rm O5}_1} = 0 \ , \nonumber 
\eeqn
where $N_{\rm O9}$ denotes the total amount of 9-brane charge carried by O9-planes and $N_{{\rm O5}_m}$ the 
5-brane charge referring to O5-planes wrapped on the two-torus $\mathbb{T}^2_m$ 
\cite{Blumenhagen:2000wh,Blumenhagen:2000ea,Cvetic:2001nr}.  
The angle variables of the dual IIA picture are defined by 
\be \label{defangle}
\varphi^{(m)}_a = {\rm arctan} \left( \frac{{\bf F}^{(m)}_a}{\Re(T_m)} \right) \ , 
\ee
which is the angle of a given stack with respect to the coordinate axis, the location of the O6-planes. 
We shall employ conventions to choose all $\varphi_a^{(m)}$ and also the relative angles $\varphi_{ab}^{(m)}$ 
between two stacks $a$ and $b$ modulo $2\pi$ in $[0,2\pi]$. This makes a distinction of cases $n_a^{(m)} \ge 0$ and 
$n_a^{(m)} \le 0$ necessary, such that in the latter case $\varphi^{(m)}_a$ lies in $[\pi/2, 3\pi/2]$ and in the 
complement otherwise. 
The condition for preserving supersymmetry in any open string sector reads \cite{Berkooz:1996km,Ohta:1997fr}
\be \label{susy}
\varphi^{(1)}_a + \varphi^{(2)}_a + 
\varphi^{(3)}_a  = 2\pi \ . 
\ee
Since one can always redefine the angles by flipping the orientations of a brane on two two-tori simultaneously, 
shifting the two angles by $\pi$, we have adopted conventions such that the sum is always $2\pi$. 
In terms of the magnetic background fields Eq.(\ref{susy}) takes the 
form 
\be \label{susyflux}
\sum_{m=1}^3 \frac{{\bf F}_a^{(m)}}{\Re(T_m)} = \prod_{m=1}^3 \frac{{\bf F}_a^{(m)}}{\Re(T_m)} \ . 
\ee
This set of conditions actually fixes generically all three 
K\"ahler parameters $\Re(T_m)$. If some of these are already fixed by the 
asymmetric orbifold projection, as happens if the group is not $\mathbb{Z}_2\times \mathbb{Z}_2$, it is just a further 
condition that necessarily has to be met by the gauge field background for the given $\Re(T_m)$. 
Since the number of brane stacks in any model will at least be four, not counting the hidden sectors, 
and the number of moduli which may vary to satisfy Eq.(\ref{susyflux}) is only three, the system is already 
overdetermined, and thus supersymmetry in the effective action below the mass scale induced by the fluxes 
should not be considered a generic feature in orbifold models. Nevertheless interesting models can be found, 
and we shall assume Eq.(\ref{susyflux}) to be fulfilled. \\

As opposed to the conventions of 
e.g. \cite{Blumenhagen:2000wh} we have now chosen a normalization which is more adapted to the notation 
in the effective field theory where gauge fluxes are independent of the geometric moduli, since they are 
quantized through the Dirac quantization (see \cite{Hashimoto:1997gm} as a useful reference).  
The integers now have an interpretation of ${\bf n}_a^{(m)}$ denoting the winding number of the brane stack 
$a$ on the torus $\mathbb{T}^2_m$ and ${\bf m}_a^{(m)}$ denoting the first Chern number of its $U(1)_a$ world volume 
gauge bundle. For $b^{(m)}=0$ 
the limiting case $({\bf n}_a^{(m)},{\bf m}_a^{(m)})=(1,0)$ has $\varphi^{(m)}_a = 0$ and describes a ``pure'' 
D-brane wrapping $\mathbb{T}^2_m$ once and without flux, 
while the degenerate case $({\bf n}_a^{(m)},{\bf m}_a^{(m)})=(0,1)$ with  $\varphi^{(m)}_a = \pi/2$ 
maps to a D-brane with Dirichlet boundary conditions on $\mathbb{T}^2_m$ after T-duality, i.e. a point like brane of 
lower dimension. This case needs actually some extra care, since ${\bf F}^{(m)}_a$ diverges formally. \\

Since the Dirac-Born-Infeld (DBI) action is in its non-abelian version only reliably known to leading order in the 
gauge field strength, we shall actually have to refer to the limit of weak fields eventually, which we define 
as ${\cal F} \rightarrow 0$. More generally, we would favor to consider a general perturbation around an exactly 
solvable orientifold background with not only D9-branes with ${\cal F}_a=0$ but also with D5-branes with some 
components of ${\cal F}_a$ being infinite, 
or even more generic rational values of ${\bf F}_a^{(m)}$ values as arise in the supersymmetric orientifolds constructed e.g. 
in \cite{Blumenhagen:1999ev,Pradisi:1999ii}. This does however not appear to be feasible. 


\subsection{Spectrum and field theory conventions} 

The total massless spectrum consists of untwisted closed string fields, 
that include the gravity plus untwisted moduli sector, of twisted closed string fields, localized at fixed points 
of $\Theta_i$, which we are mostly going to ignore throughout this paper, and finally out of open string fields. 
These again split naturally into states with both ends of the string on the same stack of brane, 
the gauge vector multiplets plus some 
extra matter in adjoint, symmetric and anti-symmetric representations, as well as those connecting different brane stacks 
in bifundamental representations, which are supposed to involve all the fermionic matter fields of the MSSM  
and the standard bifundamental Higgs fields. The former symmetric and anti-symmetric representations 
come in three copies as remnants of a ``would-be'' ${\cal N}=4$ supersymmetric theory, which arises in this sector 
upon toroidal compactification without orbifold projection, i.e. there is one complex scalar for each $\mathbb{T}^2_m$. 
We then also denote these as $C_m^{[aa]}$. For an open string between two branes at some relative angles, the 
modings in the Fourier expansion of all world sheet fields are shifted by the relative angle $\varphi_{ab}^{(m)}$ and thus 
also the zero-point energy of the NS sector. The ground state energy 
becomes (for angles smaller than $\pi/2$, else see \cite{Blumenhagen:2001te})
\be \label{zeroenergy}
E_0 = \frac12 \sum_{m=1}^3 \varphi^{(m)}_{ab} - {\rm max}\ \{ 
 \varphi^{(1)}_{ab},\varphi^{(2)}_{ab},\varphi^{(3)}_{ab} \} \ . 
\ee
For angles that satisfy Eq.(\ref{susy}) exactly one of the three components stays massless, denoted $C^{[ab]}$, 
while the other two get masses 
of the order of the mass scale associated to the fluxes through a D-term in the effective action \cite{Cremades:2002te}. 
For a more complete derivation and tables that allow to compute the precise 
spectrum from the winding numbers we again refer to the literature, for example  
\cite{Cvetic:2001nr,Blumenhagen:2002wn,Blumenhagen:2002gw,Honecker:2003vq}. \\ 

For the effective ${\cal N}=1$ field theory we use the following conventions and notations: 
The set of all fields ${\cal T}_I$, moduli and matter, 
is split into four-dimensional dilaton field $s$, 
the closed string K\"ahler and complex structure moduli 
fields $t_m$ and $u_m$, and the open string fields are denoted 
$C_\alpha=C^A_\alpha \lambda^A$, together ${\cal T}_I \in \{ s,t_m, u_m, C_\alpha\}$. Just to repeat, 
there are generically 3 $u_m$ and 0, 1, or 3 $t_m$ before applying  Eq.(\ref{susy}). 
The $\lambda^A$ span a basis of the respective representation of the gauge group $H$. 
The $C_\alpha^A$ split into $C_m^{[aa]}$ fields, which, as mentioned, transform under various representations 
of the gauge group \cite{Cvetic:2001nr} and are often ignored, and $C^{[ab]}$, transforming as bifundamental fields and 
represent quark and lepton, as well as Higgs field multiplets.  
All gauge singlets constructed out the 
$C_\alpha$ then imply traces over gauge indices, for instance  
$|C_\alpha|^2 = {\rm Tr}\, C_\alpha \bar C_\alpha = C_\alpha^A \bar C_\alpha^A$, 
${\rm Tr}\, C_\alpha C_\beta C_\gamma = \frac{i}{2} f^{ABC} C_\alpha^A C_\beta^B C^C_\gamma$. 
The auxiliary fields in the 
respective chiral multiplets are denoted $F^I\in \{ F^s, F^{t_m}, F^{u_m}\}$. 
The relation between the string frame moduli parameters $T_m,U_m$ and the ten-dimensional dilaton $\Phi$ 
and the fields in the effective four-dimensional Einstein-frame 
Lagrangian is \cite{Antoniadis:1996vw,Ibanez:1998rf} 
\be 
t_m + \bar t_{\bar m} = e^{-\Phi} ( T_m + \bar T_{\bar m} ) \ , \quad 
u_m + \bar u_{\bar m} =  U_m + \bar U_{\bar m}  \ , \quad 
s + \bar s = e^{-\Phi} \prod_{m=1}^3 ( T_m + \bar T_{\bar m} ) \ .  
\ee
The axionic fields for the imaginary parts of the fields $\{s, t_m\}$ are provided by suitable components of RR-forms 
\cite{Cremades:2002te}. Concretely, $a_0$, defined via $da_0 = *_4dC_2$, $*_4$ denoting the 
four-dimensional Hodge operator, is the partner of $s+\bar s$ and the other axions are given by certain components of $C_6$. 
We split $C_6 = C_2^{(m)} \wedge C_4^{(m)}$, where $C_4^{(m)}$ has components only along the two 2-tori transverse 
to $\mathbb{T}^2_m$, i.e. $C_4^{(m)} \propto dz_n \wedge d\bar{z}_{\bar{n}} \wedge dz_p \wedge d\bar{z}_{\bar{p}}$ with 
$n,p \not= m$. Then $da_m = *_4 dC_2^{(m)}$ defines the partner $a_m$ of $t_m+\bar t_{\bar m}$ \cite{Cremades:2002te}. 
The tree-level superpotential can be written  
\be \label{supotree}
W_{\rm tree}(u_m,C_\alpha) = \frac16 Y_{\alpha\beta\gamma}(u_m) C_\alpha C_\beta C_\gamma + \ \cdots \ 
\ee
and on general grounds only depends holomorphically on complex structure moduli $u_m$ in perturbative  
type IIB string theory \cite{Dine:1986zy,Dine:bq,Kachru:2000ih,Kachru:2000an}. 
The parenthesis indicate higher order terms. Additional contributions appear in the process of 
supersymmetry breaking in the hidden sector, such as the effective superpotential that arises by 
gaugino condensation. 
Integrating out the undetermined hidden sector a new effective superpotential can be formulated that 
involves corrections to $W_{\rm tree}$. 
As an example, gaugino condensation leads to an effective 
superpotential that depends on the moduli through the gauge kinetic functions $f_a$, that turn out to depend 
holomorphically on $s,t_m$ in our case. The precise form of the dependence on the matter fields relies on 
the structure of the $W_{\rm tree}$ above. It is obtained by setting the $C_\alpha$ in 
Eq.(\ref{supotree}) that belong to the hidden sector to their moduli-dependent vacuum expectation values. Thus the 
trilinear term in the visible $C_\alpha$ should remain unmodified, and terms with only hidden fields will generate 
a contribution without visible matter fields. The appearance of the quadratic $\mu$-term - in this approximation - 
depends on the fact, if there are fields charged under both, the hidden and the visible gauge groups, such that terms 
bilinear in visible $C_\alpha$ can be generated. In an idealized version of our model, we would want to avoid such 
fields, and disentangle the hidden and visible brane stacks on the internal space, such that they communicate 
only gravitationally. In practice, it has however not been possible to find such a model, 
and we keep the $\mu$-term to be generic. Note, however, that we would favor a situation without it. 
Together we write 
\be \label{supoeff}
W_{\rm eff} ({\cal T}_I) = W_{\rm tree}(u_m,C_\alpha) + \hat W({\cal T}_I) 
  + \frac12 \mu_{\alpha\beta}({\cal T}_I) C_\alpha C_\beta + 
\ \cdots \ . 
\ee
By a slight abuse of notation, the $C_\alpha$ now only refer to the surviving visible matter fields. 
The K\"ahlerpotential is expanded 
\be \label{kaehlerexp}
{\cal K}({\cal T}_I+\bar{\cal T}_{\bar I}) = \hat {\cal K}({\cal T}_I+\bar{\cal T}_{\bar I}) 
  + \tilde {\cal K}_{\alpha\bar \beta}({\cal T}_I+\bar{\cal T}_{\bar I}) C_\alpha \bar C_{\bar \beta} 
  + \tilde {\cal Z}_{\alpha\beta}({\cal T}_I+\bar{\cal T}_{\bar I}) C_\alpha C_{\beta} 
  + \ \cdots 
\ee
and the (holomorphic) gauge kinetic functions are written $f_a = f_a({\cal T}_I)$, $a$ labeling the factors of the gauge group. 
D-terms $D = D({\cal T}_m+\bar {\cal T}_m, C_\alpha, \bar C_{\bar \alpha})$ 
and Fayet-Iliopolous (FI) parameters, denoted $\xi_a(t_m+\bar t_m)$, may also occur. 
They depend only on the real parts of the K\"ahler moduli, due to Peccei-Quinn shift symmetries in the imaginary parts 
\cite{Dine:1986zy,Dine:bq,Kachru:2000ih,Kachru:2000an}. 
We shall always be working at leading order in the matter fields $C_\alpha$. The factorizable 
structure of the moduli space of the background torus implies that 
$\tilde {\cal K}_{\alpha\bar \beta} = \tilde {\cal K}_{\alpha} \delta_{\alpha\bar \beta}$. 
The covariant derivative with respect to ${\cal K}$ is $D=\partial + (\partial {\cal K})$ and 
the K\"ahler metric is ${\cal K}_{I\bar J} = \partial_I \partial_{\bar J}{\cal K}$.   
The auxiliary fields take values $F_I \propto D_I \hat{W}$ and the scalar potential is
 given by the standard formula \cite{Chamseddine:jx,cremmer}
\be 
V({\cal T}_I,\bar{\cal T}_{\bar I}) = e^{\cal K} 
   \left( {\cal K}^{I\bar J} D_I W_{\rm eff} \bar D_{\bar J} \bar{W}_{\rm eff} - 3|W_{\rm eff}|^2 \right) \ . 
\ee
The D-terms are assumed to vanish identically in the effective theory. 
In the vacuum only $\hat W$ will contribute and the value of $V$ is denoted by $V_0$. The gravitino mass is 
\be 
M_{3/2} = e^{{\cal K}/2} |\hat W| \ . 
\ee
Gaugino masses are given by 
\be  \label{gaugmass}
M_a = \frac{1}{2\Re (f_a)}  F^I \partial_I f_a \ . 
\ee
The soft breaking parameters are given through the effective Lagrangian 
\beqn
{\cal L}_{\rm soft} &=&  
 - m^2_{\alpha} C'_\alpha \bar C_{\bar \alpha}' 
 - \frac16 A^0_{\alpha\beta\gamma} Y^0_{\alpha\beta\gamma}  C'_\alpha  C'_\beta C'_\gamma \non 
&&  - \frac12 \left( B^0_{\alpha \beta} \mu^0_{\alpha \beta} C'_\alpha  C'_\beta + \ {\rm h.c.} \right) + \ \cdots \ . 
\eeqn
The primes and upper indices 0 indicate that the matter fields have been canonically normalized in their kinetic terms and 
suitable normalization functions been absorbed into $\mu_{\alpha\beta}$ and $Y_{\alpha\beta\gamma}$. 
The desired soft parameters $m^2_\alpha, A^0_{\alpha\beta\gamma}, B^0_{\alpha \beta}$ are the functions that multiply the 
Yukawa-couplings $Y^0_{\alpha\beta\gamma}$ and the $\mu$-parameter $\mu_{\alpha \beta}^0$. They are defined by \cite{Brignole:1997dp} 
\beqn \label{softpara}
m^2_\alpha &=& M_{3/2}^2 + V_0 - F^I \bar F^{\bar J} \partial_I \partial_{\bar J} \ln (\tilde {\cal K}_\alpha) \ , \non   
A^0_{\alpha\beta\gamma} &=& \bar c F^I \left( \partial_I \hat {\cal K} + \partial_I \ln (Y_{\alpha\beta\gamma}) 
  - \partial_I \ln ( \tilde {\cal K}_\alpha \tilde {\cal K}_\beta \tilde {\cal K}_\gamma ) \right)  \ , \non 
B^0_{\alpha\beta} &=& \bar c F^I \left( \partial_I \hat {\cal K} + \partial_I \ln(\mu_{\alpha\beta}) - 
 \partial_I \ln (\tilde {\cal K}_{\alpha} \tilde {\cal K}_{\beta} )  \right) + \ \cdots \ , 
\eeqn
where $\bar c$ is some normalization that will be put in later.
In leaving out additional terms in the third line of Eq.(\ref{softpara}) we have assumed that the term $\tilde{\cal Z}_{\alpha\beta}$ 
in the K\"ahler  potential vanishes, or that and all such terms are transferred to 
the superpotential by a K\"ahler transformation. 


\section{Elements of the effective action}

The ingredients to perform explicit calculations are obviously the functions that determine the effective Lagrangian. 
In order to obtain expressions for these, we have translated the intersecting brane world scenario back into 
the type I language of the IIB picture, 
which allows us to use dimensional reduction of standard ten-dimensional type I expressions, 
and further employ the (partly) perturbative duality to the heterotic string, whose effective action 
is expected to be very similar in many 
respects \cite{Chamseddine:ez,Witten:1985xb,Burgess:1985zz,Ferrara:1986qn}. 
This would not be easily possible in the type IIA picture with intersecting 
D6-branes and O6-planes. To keep formulas handy we specialize to the case of the $\mathbb{Z}_2 \times \mathbb{Z}_2$ orbifold 
group, i.e. we always keep 3 $t_m$ moduli. \\ 

\subsection{$\hat {\cal K}$, $\tilde{\cal K}_{\alpha\bar \beta}^{[aa]}$ and $f_a$}

Before getting started let us actually cite previous expressions given for the type I K\"ahler potential 
in the presence of D9- and D5-branes. In \cite{Ibanez:1998rf} such were given for the case of all 
tori being squares, $U_m = 1$, and $B_2=0$ such that T-duality can be used very simply to infer the terms for the 
three possible types of supersymmetry preserving D5-branes wrapped on 
any single $\mathbb{T}^2_m$ from those 
of D9-branes, the latter being taken from the perturbative heterotic result by use of \cite{Antoniadis:1996vw}. 
The open string fields are now denoted $C_m^{[99]}$ and $C_m^{[5_n 5_n]}$ for the $m$-component of the massless bosonic 
excitation of an open string with both ends on a given 9-brane, or on a D5$_n$-brane wrapping $\mathbb{T}^2_n$.  
Analoguously, one has $C^{[5_m 5_n]}$ and $C^{[95_m]}$ as the massless bosonic NS-ground state  of an open string connecting 
two different D5-branes wrapped on two different tori $m$ and $n$ or connecting a 9-brane and a 5-brane. 
These branes are degenerate examples of magnetic fluxes corresponding to
\beqn \label{d9d5flux}
{\rm D9} &:& \quad ( {\bf n}_a^{(m)} , {\bf m}_a^{(m)} ) = (\pm 1,0) \ {\rm for\ all\ }m \ , \non  
{\rm D5}_n &:& \quad ( {\bf n}_a^{(n)} , {\bf m}_a^{(n)} ) = (\pm 1,0) \ , 
   \ ( {\bf n}_a^{(p)} , {\bf m}_a^{(p)} ) = (0,\pm 1)\ {\rm for\ } p\not= n \ . 
\eeqn
The branes have to be defined with appropriate orientation so that the relative angles work out to fulfill Eq.(\ref{susy}). 
The important point now is that a combined T-duality along two among the three 2-tori, say all except $\mathbb{T}^2_n$,  
is a symmetry of the effective action, by exchanging D9 $\leftrightarrow$ D5$_n$ and $s \leftrightarrow t_n$, 
$t_m \leftrightarrow t_p$ for $m\not= n \not= p$. The full T-duality  
group of the $\mathbb{T}^6$ will actually no longer be a symmetry of the 
background. \\ 
 
Let us first concentrate on the fields 
$C^{[aa]}_m$, i.e. only regard $C_m^{[99]}$ and $C_m^{[5_n 5_n]}$. Their metric was written 
\beqn \label{ibanez1}
\hat {\cal K} 
&=& 
-\ln( s+ \bar s) - \sum_{m=1}^3 \ln( t_m + \bar t_{\bar m} ) \ , \\ 
\frac12 \tilde{\cal K}_{\alpha\bar \beta}^{[aa]} C^{[aa]}_{\alpha} \bar C^{[aa]}_{\bar\beta} 
&=&
\sum_{m=1}^3 \frac{|C_m^{[99]}|^2}{t_m + \bar t_{\bar m}} 
  + \sum_{m=1}^3 \frac{|C_m^{[5_m 5_m]}|^2}{s + \bar s} 
  + \frac12 \sum_{m,n,p=1}^3 \gamma_{mnp} \frac{|C_n^{[5_m 5_m]}|^2}{t_p + \bar t_{\bar p}} \ , 
\nonumber
\eeqn
using $\gamma_{mnp} = 1$ for $m\not= n \not= p \not= m$ and 0 else. 
Since the magnetic field background only affects the open string fields, the first line of Eq.(\ref{ibanez1})  
is the standard K\"ahler 
potential for 
\be 
\left( \frac{SU(1,1)}{U(1)} \right)_s \times \left( \frac{SU(1,1)}{U(1)} \right)^3_{t} \ , 
\ee
the scalar manifold of 
the untwisted moduli of a toroidal orbifold without $u_m$ moduli. It remains unchanged and also applies to our models. 
The first term in the second line of Eq.(\ref{ibanez1})  
is identical to the heterotic potential for an untwisted matter field, 
corresponding to 
\be \label{mani2}  
\left( \frac{SU(1,1)}{U(1)} \right)^3_{t} ~\rightarrow~ \left( \frac{SU(1,1+N)}{U(1)} \right)^3_{t} \ , 
\ee
and the other two follow from applying T-dualities along any two among the three two-tori. 
Here $N$ is the number of extra matter fields $C_m$. 
The role of D5-branes in type I vacua is unconventional if 
compared to the heterotic string, and partly appears as a non-perturbative effect there. In particular, in the heterotic 
K\"ahler potential the dilaton $s+\bar s$ does not appear in the matter metric. 
The above expressions will later be reproduced as special cases 
by our more general formulas upon applying Eq.(\ref{d9d5flux}). \\ 

We now turn to computing the K\"ahler metric for the $C_\alpha^{[aa]}$ strings from a simple direct reduction of the 
(abelian) Dirac-Born-Infeld (DBI) effective Lagrangian. Since we are dealing with strings with both ends on the same brane 
we do not really have to invoke the full (unknown) non-abelian version of the DBI action, but can restrict ourselves 
to a single stack of a single brane, if necessary. Nevertheless we keep the formulas non-abelian. 
The general form for a D$p$-brane labelled by $a$ is 
\be \label{dbi} 
{\cal S}_{\rm DBI} = - \mu_p \ {\rm Tr}\ \int_{{\cal W}_a^{p+1}} d^{p+1}\xi\ e^{-\Phi}
\sqrt{ - {\rm det} \left( {\rm P}[ G+B ] + F_a \right) }  \ . 
\ee
The same formula measures the tension of the orientifold planes, setting the gauge fields to zero. 
The pull-back to the world volume ${\cal W}_a^{p+1}$ 
is trivial for $p=9$, and world volume and space-time fields and coordinates can be identified. 
For the abelian case we can also just drop the trace, which contains all the non-abelian information and problems. 
One only needs to take account of the multiple wrapping of the branes when writing the integral over the world volume 
as an integral over the torus. For this reason we rescale the background fields and introduce 
${\bf n}_a\cdot G$ and ${\bf n}_a\cdot {\cal F}_a$ by 
\be
{\bf n}_a\cdot G = {\rm diag} ( {\bf n}_a^{(1)}G^{(1)},{\bf n}_a^{(2)}G^{(2)},{\bf n}_a^{(3)}G^{(3)} ) \ , \quad {\rm  etc.}
\ee
The scalar matter fields $C_\alpha^{[aa]}$ now arise as internal components of the world volume gauge fields 
$A_{a}$\footnote{The upper capital gauge group indices will be suppressed 
most of the time, while the lower index $a$ for the brane stack is written. This is a redundant labeling anyway, since 
the stacks are simultaneously counted as factors of the gauge group.} inside 
$F_a$, which split into four-dimensional scalars $A_{ai}$ and vectors $A_{a\mu}$. 
>From  their kinetic terms we want to read off the moduli space metric.  
We combine $F_a+B=\tilde{\cal F}_a$ and write $\tilde{\cal F}_a = 
{\cal F}_a + \delta {\cal F}_a$, where 
\beqn
(\delta {\cal F}_a)^A_{\mu\nu} &=& 2 \partial_{[\mu } A^A_{\nu ]a} + f^{ABC} A^{B}_{a\mu} A^{C}_{a\nu} \ , \non 
(\delta {\cal F}_a)^A_{\mu i} &=& D_\mu A^A_{ai} ~=~ \partial_\mu A^A_{ai} + f^{ABC} A^{B}_{a\mu} A^C_{ai} \ , \non 
(\delta {\cal F}_a)^A_{ij} &=& f^{ABC} A^{B}_{ai} A^{C}_{aj} 
\eeqn
(now letting $i,j= 1,\, ...\, , 6$) contains only the fluctuations of the four-dimensional vector fields $A_{a\mu}$, 
and scalars $A_{ai}$. 
${\cal F}_a$ is the constant background in the Cartan subalgebra, or more precisly 
in the $U(1)_a$ subgroup explicit in $U(N_a)= (SU(N_a) \times U(1)_a)/\mathbb{Z}_N$, such that $SU(N_a)$ 
remains unbroken. 
To extract the leading terms of the DBI Lagrangian in the fluctuations one uses 
\beqn \label{dettr}
\sqrt{{\rm det}(1+M)} = 1 + \frac12 {\rm tr} (M) + \frac18 \left( {\rm tr} (M) \right)^2 
 - \frac14 {\rm tr}( M^2) + \ \cdots \ . 
\eeqn
Here the trace only refers to Lorentz indices. 
The kinetic terms then read 
\beqn
{\cal S}_{\rm DBI} &=& - \mu_9 \int d^{10}x\ \sqrt{-g_4} 
 \sum_a \sqrt{ {\rm det}( {\bf n}_a\cdot G+{\bf n}_a\cdot {\cal F}_a) } e^{-\Phi} \\ 
&& \hspace{0.5cm} 
 \times \left( \frac12 ( {\bf n}_a\cdot G + {\bf n}_a\cdot {\cal F}_a )^{ij} g^{\mu\nu} D_\mu A_{ai} D_\nu A_{aj} 
+ \frac14 (\delta {\cal F}_a)_{\mu\nu}(\delta {\cal F}_a)^{\mu\nu} \right) \ + \ \cdots 
\nonumber
\eeqn
We use the convention that $(G+{\cal F}_a)^{ij}$ is the inverse of $(G+{\cal F}_a)_{ij}$.  
The gauge coupling $g_{10}$ enters via 
\be
\mu_9 = \frac{1}{g^2_{10}} ( 2\pi \alpha')^{-2} \ , 
\ee
where the extra factors $2\pi \alpha'$ are absorbed by rescaling the fields, and are set to 1 in our 
conventions anyway. 
The gauge kinetic functions are defined by the prefactor in 
\beqn
-\frac{\Re(f_a)}{4 g_{10}^2} (\delta {\cal F}_a)_{\mu\nu}(\delta {\cal F}_a)^{\mu\nu} \ , 
\eeqn 
and using the supersymmetry condition Eq.(\ref{susy}) one can show that 
\beqn \label{vola} 
\Re (f_a) &=& 
e^{-\Phi} \sqrt{ {\rm det}( {\bf n}_a\cdot G+{\bf n}_a\cdot {\cal F}_a) } \\ 
&=& 
\frac12 {\bf n}_a^{(1)}{\bf n}_a^{(2)}{\bf n}_a^{(3)} (s+\bar s) 
  - \frac14 \sum_{m,n,p=1}^3 \gamma_{mnp} {\bf n}_a^{(m)} {\bf m}_a^{(n)} {\bf m}_a^{(p)} (t_m + \bar t_{\bar m}) \ . 
\nonumber
\eeqn
Under a Weyl-rescaling of the four-dimensional metric $g_{\mu\nu}$ the kinetic term for the gauge field 
fluctuations $A_{a\mu}$ is invariant, such that Eq.(\ref{vola}) is the answer in the Einstein frame.  
Together with the Chern-Simons (CS) action 
\beqn \label{csaction}
{\cal S}_{\rm CS} = \mu_9 \sum_a \int_{{\cal W}_a^{10}} d^{10}\xi\  
 e^{-{\cal F}_a} \wedge \left( C_2 + C_6 \right)  \ , 
\eeqn
that contains the axionic couplings of the imaginary parts in $C_2$ and $C_6$, the real part combines into the 
gauge kinetic function \cite{Cremades:2002te}
\be
f_a(s,t_m) = {\bf n}_a^{(1)}{\bf n}_a^{(2)}{\bf n}_a^{(3)} s
  - \frac12 \sum_{m,n,p=1}^3 \gamma_{mnp} {\bf n}_a^{(m)} {\bf m}_a^{(n)} {\bf m}_a^{(p)} t_m  \ . 
\ee
In \cite{Lust:2003ky} the threshold corrections to gauge couplings have been calculated as well.  
The matter metric for the $A_{ai}$ transforms under the Weyl-rescaling given by 
$g_{\mu\nu} \mapsto e^{2\Phi} \sqrt{G}^{-1} g_{\mu\nu}$, 
that puts the gravitational Lagrangian into standard Einstein-Hilbert form and one gets
\be \label{weyl}
e^{\Phi} \sqrt{ \frac{{\rm det}( {\bf n}_a\cdot G+{\bf n}_a\cdot {\cal F}_a) }{{\rm det}(G)}} 
  ( {\bf n}_a\cdot G + {\bf n}_a\cdot {\cal F}_a )_{\rm sym}^{ij} ~=~
\frac{2 \Re (f_a)}{s+\bar s} e^\Phi ( {\bf n}_a\cdot G + {\bf n}_a\cdot {\cal F}_a )_{\rm sym}^{ij}
\ee
for the proper matter metric. Using the factorization of the six-torus we specialize to a single two-torus 
\beqn \label{metric1}
e^\Phi \left( {\bf n}_a^{(m)} G^{(m)} + {\bf n}_a^{(m)} {\cal F}^{(m)}_a \right)_{\rm sym}^{-1} ~=~
\frac{1}{1 + \Delta_a^{(m)}} \left( e^{-\Phi} {\bf n}_a^{(m)} G^{(m)} \right)^{-1} \ ,
\eeqn
having defined 
\beqn
\Delta_a^{(m|np)} ~=~
\frac{ ( t_n +\bar t_{\bar n} )( t_p +\bar t_{\bar p} )}
  {(s+\bar s)( t_m +\bar t_{\bar m} )} 
\left( {\bf F}_a^{(m)} \right)^2 \ , \quad 
\Delta^{(m)}_a ~=~ \frac12 \sum_{n,p=1}^3 \gamma_{mnp} \Delta_a^{(m|np)}  \ . 
\eeqn
The significance of this $\Delta_a^{(m)}$ is that it effectively summarizes explicit moduli dependence 
that will appear in the soft breaking parameters. We see that the K\"ahler metric for any two-torus 
is equal to the metric on the manifold Eq.(\ref{mani2}) only rescaled by a factors that depends on ${\cal F}_a$. \\ 

For the sake of comparing to Eq.(\ref{ibanez1}) we specify to the simple case where all $U_m = 1$ and $b^{(m)}=0$, such that 
$2 e^{-\Phi}{\bf n}^{(m)}_a G^{(m)}_{ij} = {\bf n}^{(m)}_a {\rm diag}(t_m+\bar t_{\bar m},t_m+\bar t_{\bar m})_{ij}$. 
In this limit we shall identify 
the $A_{a2m-1}+iA_{a2m}$ with the K\"ahler coordinates $C_m^{[aa]}$. 
The D9-brane case, $({\bf n}_a^{(m)},{\bf m}_a^{(m)})=( 1,0)$ for all $m$, then easily reproduces the result 
of Eq.(\ref{ibanez1}) 
for the $C_m^{[99]}$ fields, since $2 \Re (f_a) = {s+\bar s}$ and $\Delta^{(m)}_a = 0$. 
Only slightly more challenging are the components of world volume gauge fields along 5-branes, 
the $C_m^{[5_m5_m]}$ fields, which are reproduced upon choosing 
$({\bf n}_a^{(m)},{\bf m}_a^{(m)})=( 1,0)$ and $({\bf n}_a^{(p)},{\bf m}_a^{(p)})=(0, 1)$ 
along the other two two-tori with $p \not= m$. 
Note that then $2 \Re (f_a) = t_m +\bar t_{\bar m}$ and again $\Delta^{(m)}_a = 0$. Thus, the metric 
Eq.(\ref{metric1}) is 
a deformation of the K\"ahler metric on $SU(1,1+N)/U(1)$ and reduces to the standard metric upon switching off the flux, 
but also knows about the heterotically non-perturbative 5-brane sectors. As another comparison, one may easily 
recognize the so-called ``open string metric'' of \cite{Seiberg:1999vs} 
in Eq.(\ref{metric1}), which guarantees that it describes 
the correct KK mass spectrum of massive excitations in the gauge field background \cite{Blumenhagen:2000fp}, 
or equivalently, on a non-commutative torus. That the string spectrum 
coincides with the field theory approximation \cite{vanBaal:1984ar,Troost:1999xn} 
has been shown in \cite{Hashimoto:1997gm,Denef:2000rj}. \\ 

The metric for the transverse scalars of a D5-brane, the $C_n^{[5_m5_m]}$ fields, needs to be treated separately. 
It would be described by choosing $({\bf n}_a^{(m)},{\bf m}_a^{(m)})=( 1,0)$ and 
$({\bf n}_a^{(p)},{\bf m}_a^{(p)})=(0, 1)$ for $p\not= m$ as 
before, but then Eq.(\ref{metric1}) vanishes. This is, however, not surprising since the fluctuations of the transverse 
scalars, denoted $A^i_a$, of any lower-dimensional brane are not internal components of world volume gauge fields, 
but enter the DBI action via the non-trivial (and non-abelian \cite{Myers:1999ps}) 
pull-back, once the brane is not ten-dimensional. By substituting 
\be
{\rm P}[g_{\mu\nu}] = g_{\mu\nu} + G_{kl} D_\mu A^{k} D_\nu A^l + \ \cdots 
\ee
and ${\cal F}^{(m)}_a = 0$ into Eq.(\ref{dbi}), one finds the kinetic term  
\beqn
{\cal S}_{\rm DBI} &=& - \mu_9 \int d^{10}x\ \sqrt{-g_4} 
 \sqrt{ {\rm det}( G ) } e^{-\Phi} 
G_{ij} g^{\mu\nu} D_\mu A_{a}^i D_\nu A_{a}^j \ + \ \cdots 
\eeqn
Here it is important that the $A_a^i = G^{ij} A_{aj}$ are used as independent 
K\"ahler coordinates, not the $A_{ai}$ (see e.g. \cite{Myers:1999ps}). Therefore, the matter metric is  
\be
\frac{2\Re (f_a)}{s+\bar s} e^{2\Phi} \left( e^{-\Phi} G^{(m)}_{ij} \right) = 
\frac{8\Re (f_a)}{\prod_{p=1}^3(t_p + \bar t_{\bar p})}  \left( e^{-\Phi} G^{(m)}_{ij} \right) \ . 
\ee
For the scalars $C^{[5_m5_m]}_n$ we have to use $({\bf n}_a^{(m)},{\bf m}_a^{(m)})=( 1,0)$ 
and $({\bf n}_a^{(p)},{\bf m}_a^{(p)})=(0, 1)$ for $p\not= m$ 
as mentioned above, and via $2e^{-\Phi} G^{(n)}_{ij} = {\rm diag} (t_n + \bar t_{\bar n},t_n + \bar t_{\bar n})$ and 
$2\Re (f_a) = t_m + \bar t_{\bar m}$ just obtain the third term of Eq.(\ref{ibanez1}). The same procedure would apply 
to the transverse scalars of D7-branes or D3-branes as well. \\ 

For the general dependence of the K\"ahler metric in the presence of $U_m$ moduli as well, we just note that 
\beqn 
e^{\Phi} \left( G^{(m)}\right)^{ij} D_\mu A_i D^\mu A_j &=& \frac{4}{(t_m + \bar t_{\bar m})(u_m + \bar u_{\bar m})} 
  | u_m D_\mu A_{2m-1} + i D_\mu A_{2m} |^2 \ , \non  
e^{-\Phi} \left( G^{(m)}\right)_{ij} D_\mu A^i D^\mu A^j &=& \frac{t_m + \bar t_{\bar m}}{u_m + \bar u_{\bar m}} 
  | i D_\mu A^{2m-1} - u_m D_\mu A^{2m} |^2 \ . 
\eeqn 
Therefore, the correct K\"ahler coordinates for the open string fields are defined by
\cite{LopesCardoso:1994is,Antoniadis:1996vw} 
\be \label{opkae}
C_m^{[aa]} =  u_m A_{a2m-1} + i A_{a2m} \  \quad {\rm or}\quad  
C_m^{[aa]} =  i A_a^{2m-1} - u_m A_a^{2m} \ , 
\ee
for the two longitudinal components $A_i$ of world volume vectors or the two transverse scalars $A^i$ along any 
$\mathbb{T}^2_m$, respectively. This reproduces the correct K\"ahler metric for the moduli space 
\be
\left( \frac{SU(1,1)}{U(1)} \right)_s \times \left( \frac{SO(2,2+N)}{SO(2)\times SO(2+N)} \right)^3_{t,u} \ , 
\ee
which is known to be correct for the orbifolds of the heterotic string \cite{LopesCardoso:1994is}. 
The effect of the gauge flux is merely a moduli-dependent rescaling of this metric. While the K\"ahler potential for the 
$SU(1,1)/U(1)$ moduli space, ${\cal K} = - \ln(t_m +\bar{t}_{\bar{m}} - |C^{[aa]}_m|^2 ) +\ \cdots$ does not induce 
holomorphic or anti-holomorphic terms like $C^{[aa]}_m C^{[aa]}_m$ in the effective action, the K\"ahler potential 
of the more general case $SO(2,3)/(SO(2)\times SO(3))$ \cite{LopesCardoso:1994is}
\be \label{socoset}
{\cal K} = - \ln \left( (t_m +\bar{t}_{\bar{m}})(u_m +\bar{u}_{\bar{m}}) 
  - \frac12 ( C^{[aa]}_m + \bar C^{[aa]}_{\bar m} )^2 \right) + \ \cdots  
\ee 
does. In other words, already in the model undeformed by magnetic flux the coefficients 
$\tilde{\cal Z}_{\alpha\beta}$ in Eq.(\ref{kaehlerexp}) are non-vanishing (even equal to the coeffcicients $\tilde{\cal K}_{\alpha\beta}$), 
and thus we expect that generically these will survive 
in the deformed theory. To compute their contribution in the effective potential in the same way as the metric 
above from a dimensional reduction we would have to use the non-abelian DBI action. For our present purposes, 
these coefficients are not so important, and we set them to zero henceforth. They would be relevant for discussing 
the $\mu$-problem, of course, as in \cite{Brignole:1997dp}. \\ 

Since the case of Dirichlet boundary conditions for D5- or even D7- or D3-branes is somewhat more special and one 
can always adapt to it easily, we henceforth only regard branes with mixed Neumann-Dirichlet or pure Neumann 
boundary conditions, i.e. D9-branes with regular gauge fluxes, to simplify the notation. 
Just to collect the result of this section, the K\"ahler potential so far reads
\beqn \label{kaepot}
\hat {\cal K}({\cal T}_I + \bar{\cal T}_{\bar I} ) 
&=& 
-\ln( s+ \bar s) - \sum_{m=1}^3 \ln( t_m + \bar t_{\bar m} ) - \sum_{m=1}^3 \ln( u_m + \bar u_{\bar m} ) \ , \\ 
\tilde{\cal K}_{m\bar m}^{[aa]} ({\cal T}_I + \bar{\cal T}_{\bar I} )  
&=&
\frac{1}{(s+ \bar s)(t_m + \bar t_{\bar m})(u_m + \bar u_{\bar m})} \frac{4\Re (f_a)}{1 + \Delta_a^{(m)}}   
\nonumber   
\eeqn
with $\Re (f_a)$ given in Eq.(\ref{vola}).  
In the weak field limit ${\cal F}_a \rightarrow 0$, 
\be 
\frac{2\Re (f_a)}{1 + \Delta_a^{(m)}} = (s+\bar{s} ) + {\rm o}( {\cal F}_a) \ ,
\ee
and we arrive at the metric derived from Eq.(\ref{socoset}). While this case is well-known from the heterotic string or 
type I models with only D9-branes, the full expressions (\ref{kaepot}) is completely new and will lead to various 
novel effects when computing the soft breaking terms.


\subsection{Twisted open strings: $\tilde{\cal K}^{[ab]}_{\alpha\bar\beta}$}

We now turn to the K\"ahler potential for the open strings with ends on two different 
D-branes, two D9-branes with different gauge fields ${\cal F}_a$ and ${\cal F}_b$. In the T-dual version they 
stretch between two D6-branes at some non-vanishing relative angle  
\be \label{smallangles}
\varphi^{(m)}_{ab} = \varphi^{(m)}_a - \varphi^{(m)}_b \ .
\ee
They can be considered twisted open strings 
in the sense that the oscillator modings of the world sheet fields are shifted by rational numbers in a fashion 
very reminiscent of twisted closed string sectors in orbifold compactifications. The zero-point energy of the 
NS sector is shifted and the lowest excitation is given in Eq.(\ref{zeroenergy}). 
Therefore, when turning on a relative rotation continuously, splitting a stack $a$ into two 
stacks $b$ and $c$, two of the three massless scalars $C_{m\bar m}^{[aa]}$ become massive and only one survives as 
$C^{[bc]}$. In the effective action, the massive fields are assumed being integrated out. 
Unfortunately, in a compact background the deformation cannot be done continuously for 
flat branes, but involves discrete jumps. \\ 

In \cite{Ibanez:1998rf} the duality to the heterotic 
string \cite{Polchinski:1995df}, 
where the metric for twisted matter fields is known \cite{Cvetic:1988yw,Dixon:1989fj,Ibanez:1992hc}, 
was employed to write down a proposal for their K\"ahler potential in the presence of D9- and D5-branes. 
Since any attempt to find the metric for these $C^{[bc]}$ fields directly via a dimensional 
reduction would involve the full non-abelian DBI action, we are unable to follow the same path as in the previous 
section\footnote{It was noted in \cite{Hashimoto:1997gm,Denef:2000rj} 
that a symmetrized trace prescription, proposed in \cite{Tseytlin:1997cs}, is not entirely 
sufficient to describe open strings stretching between two branes at a relative angle, producing not quite 
the correct mass spectrum in a general case. In fact, it is believed that the order ${\cal F}^4$ is 
correctly given by the DBI action with symmetrized trace, and should then be comparable to our formulas.}
 and rely on the same duality arguments to determine the relevant terms in the effective action. 
In its probably best known example, the duality of the type I and heterotic string with gauge group $SO(32)$ compactified to 
four dimensions on a K3-orbifold space $\mathbb{T}^4/\mathbb{Z}_2 \times \mathbb{T}^2$ 
requires a matching of open strings with both ends on D9-branes 
with untwisted heterotic matter fields, and open strings between D9- and D5-branes with 
twisted fields \cite{Antoniadis:1996vw,Antoniadis:1997nz}.  
Finally, strings with both ends on a D5-brane are non-perturbative heterotic excitations, such as shrunken 
instantons, invisible in the perturbative effective action. 
The rank of the perturbative heterotic gauge group is then one half of the type I rank. 
In the simplest example of a CY-orbifold $\mathbb{T}^6/\mathbb{Z}_3$ 
one can combine knowledge about the tree level string 
spectrum and the superpotential to demonstrate the low energy theories on 
both sides may have identical massless degrees of freedom \cite{Kakushadze:1997wx,Kakushadze:1998cd}. In more 
general ${\cal N}=1$ orbifold models more intricate structures arise, it can, for example, happen that now some twisted 
heterotic matter does not have a perturbative type I origin, see \cite{Kakushadze:1998cd,Kakushadze:1999cx,Gregori:2001ak}. 
The models of interest here are even asymmetric orientifold 
vacua with reduced rank of the gauge group, whose heterotic duals are so far not known. 
The assumptions that underlie ${\cal N}=1$ heterotic-type I 
duality on CY-orbifold spaces may thus not have the status of proven facts, but we will use the analogy 
between twisted heterotic matter fields and open type I strings stretching between D-branes at relative angles, 
relying on that there is an overlap of perturbative heterotic and perturbative type I sectors. \\ 

The explicit expressions of \cite{Ibanez:1998rf} were further based on the invariance of the action under 
T-duality in a similar way as for Eq.(\ref{ibanez1}). For $U_m=1$ and $B_2=0$ this lead to  
\beqn \label{ibanez2}
\frac12 \tilde{\cal K}^{[ab]}_{\alpha\bar\beta} C^{[ab]}_\alpha \bar C^{[ab]}_{\bar \beta} &=& 
\frac12 \sum_{m,n,p=1}^3 \gamma_{mnp} \frac{|C^{[95_m]}|^2}{(( t_n +\bar t_{\bar n} )( t_p +\bar t_{\bar p}))^{1/2}} 
\non 
&&
+~ \frac12 \sum_{m,n,p=1}^3 \gamma_{mnp} \frac{|C^{[5_n5_p]}|^2}{(( s+\bar s )( t_m +\bar t_{\bar m}))^{1/2}} \ .
\eeqn
Again, the first line is the perturbative heterotic result for the three twisted sectors of a $\mathbb{Z}_2 \times 
\mathbb{Z}_2$ orbifold, upon applying 
the trivial identification of moduli fields \cite{Antoniadis:1996vw,Antoniadis:1997nz}
\beqn 
s_m^{\rm het} \mapsto s_m \ , \quad t_m^{\rm het} \mapsto t_m \ , \quad u_m^{\rm het} \mapsto u_m \ . 
\eeqn
The three terms in the first line of 
 Eq.(\ref{ibanez2}) refer to the three possible open strings between a D9-brane and wrapped D5-branes, 
the second line is obtained by T-duality along two among the three two-tori again. An important point is 
the different definition of the moduli fields in type I compared to the heterotic string: The $t_m$ depend 
on the ten-dimensional dilaton, while the $t_m^{\rm het}$ do not \cite{Antoniadis:1996vw}. \\ 

We will now apply a similar reasoning to the generalized open string sectors between branes at relative angles, 
assuming that there the K\"ahler metric can be derived by translating the tree level 
heterotic metric for a twisted sector matter field \cite{Cvetic:1988yw,Dixon:1989fj,Ibanez:1992hc}.\footnote{In a 
world sheet sense the analogy between twisted open strings and true twisted 
closed strings intuitively boils down to taking the square root of the sphere diagrams to get the results 
for discs \cite{Cvetic:2003ch,Abel:2003vv,Klebanov:2003my}.}   
\be \label{kaehettw}
\tilde{\cal K}^{{\rm het}}_{\alpha\bar\beta} =
\delta_{\alpha\bar\beta} 
 \prod_{m=1}^3 ( t^{\rm het}_m +\bar t^{\rm het}_{\bar m} )^{v^{(m)}-1}  
 \prod_{m=1}^3 ( u^{\rm het}_m +\bar u^{\rm het}_{\bar m} )^{v^{(m)}-1}  
\ .   
\ee
Generally, any generator of an orbifold group is defined by a twist vector $v=(v^{(1)},v^{(2)},v^{(3)})$ through 
its eigenvalues $\exp(2\pi i v^{(m)})$ when acting on the complex coordinates of $\mathbb{T}_m^2$, 
where we choose conventions $v^{(1)} + v^{(2)} + v^{(3)} =1$ or $2$, and $v^{(m)} \in (0,1]$.\footnote{These are 
slightly non-standard conventions as compared to $v^{(m)} \in [0,1)$ \cite{Dixon:1989fj,Ibanez:1992hc}. 
We allow $v^{(m)}=1$ to avoid the distinction of two cases 
in the final formulas. One has $v^{(m)}=1$ for at most one $m\in\{1,2,3\}$, and  $v^{(1)} + v^{(2)} + v^{(3)} =2$ 
if and only if this case, one  $v^{(m)}=1$, occurs. See equations (2.21) to (2.28) in 
\cite{Ibanez:1992hc} for a comparison.} 
A twisted NS or R world sheet oscillator field of lowest energy has then modings 
in $1/2 \pm v^{(m)} + \mathbb{Z}$ or $v^{(m)} \pm \mathbb{Z}$ respectively, and similar shifts apply to the 
coefficients in the OPE of vertex operators. 
On the other hand, an open string sector of strings stretching between two D6-branes
is similarly defined by a vector of relative (oriented) angles 
$(\varphi^{(1)}_{ab},\varphi^{(2)}_{ab},\varphi^{(3)}_{ab})$, 
$\varphi^{(m)}_{ab} \in [0,2\pi]$ and subject to the supersymmetry constraint Eq.(\ref{susy}). 
Since the twisted open strings that stretch between branes at relative angles are subject to the 
same kind of shifts in their oscillator modings and OPE, 
one would now like to just identify the relative angles with the heterotic shift 
vectors and apply formula Eq.(\ref{kaehettw}) to the twisted open strings. There are, however, some 
practical subtleties concerning the translation of the parameters and the accomodation of the non-perturbative sectors 
connected to D5-branes. \\

For the shift $\nu_{ab}^{(m)}$ in the K\"ahler potential of the open string fields  
the orientation of the branes playes a role, but the orientation of the open string does not, since  
we want the K\"ahler metric for $C^{[ab]}$ and $C^{[ba]}$ to be equal. We therefore suggest 
to use the standard formula for an oriented angle between two 
vectors $({\bf n}_a^{(m)} R_2,{\bf m}_a^{(m)}/R_1)$ in $\mathbb{R}^2$, 
\be \label{cos}
\nu^{(m)}_{ab}  = \frac{1}{\pi} {\rm arccos} \left( \frac{ 1 + {\bf F}_a^{(m)}{\bf F}_b^{(m)} \Re(T_m)^{-2}}{
                                   \prod_{c=a,b}\sqrt{ 1 + ( {\bf F}_c^{(m)} )^2 \Re(T_m)^{-2} } } \right) \ , 
\ee
to define the analogue of a $v^{(m)}$ in the heterotic string for the shift between the stack $a$ and $b$. 
This measures the angle only in $[0,\pi]$ and we have $\nu_{ab}^{(m)} = \nu_{ba}^{(m)}$. This means, we identify 
a shift vector $(\nu^{(1)}_{ab},\nu^{(2)}_{ab},\nu^{(3)}_{ab})$ with the relative angle by 
$\nu^{(m)}_{ab} = \varphi^{(m)}_{ab}/\pi$ or $2 - \varphi^{(m)}_{ab}/\pi$ depending on $\varphi^{(m)}_{ab}$ being 
less or bigger than $\pi$. 
The condition Eq.(\ref{susy}) for the angles now translates into 
\be
\nu_{ab} = \nu^{(1)}_{ab} + \nu^{(2)}_{ab} + \nu^{(3)}_{ab} \in [0,2] \ . 
\ee
Then Eq.(\ref{kaehettw}) maps naively to 
\beqn  
\tilde{\cal K}^{\rm het}_{\alpha\bar\beta} ~\mapsto~ \tilde{\cal K}^{[ab]}_{\alpha\bar\beta} &=&
\delta_{\alpha\bar\beta} 
 \prod_{m=1}^3 ( t_m +\bar t_{\bar m} )^{\nu^{(m)}_{ab}-1}  
 \prod_{m=1}^3 ( u_m +\bar u_{\bar m} )^{\nu^{(m)}_{ab}-1}  \non 
&=& 
\delta_{\alpha\bar\beta} 
e^{\Phi(3-\nu_{ab})} 
 \prod_{m=1}^3 ( T_m +\bar T_{\bar m} )^{\nu^{(m)}_{ab}-1} 
 \prod_{m=1}^3 ( U_m +\bar U_{\bar m} )^{\nu^{(m)}_{ab}-1} \ .   
\eeqn 
and the first perturbative line of Eq.(\ref{ibanez2}) is reproduced for $\nu^{(m)}_{ab} = (1/2,1/2,1)$ and 
permutations thereof. 
However, this cannot be the general answer, since the dependence on the ten-dimensional dilaton does not always 
match the proper power $\exp(\Phi)$, expected for the kinetic term that stems from a disc diagram and 
after the Weyl rescaling to the Einstein frame. Instead, the perturbative heterotic K\"ahler potential only 
leads to an acceptable perturbative K\"ahler metric in type I if the orientations are chosen such that $\nu_{ab}=2$. 
Turning the argument around, we can start with the correct dilaton prefactor in type I and rewrite  
\beqn  
e^{\Phi} 
 \prod_{m=1}^3 ( T_m +\bar T_{\bar m} )^{\nu^{(m)}_{ab}-1} 
=
(s+\bar s)^{\nu_{ab}/2-1}
 \prod_{m=1}^3 ( t_m +\bar t_{\bar m} )^{\nu^{(m)}_{ab}-\nu_{ab}/2}
\ , 
\eeqn
or 
\beqn \label{twkae} 
\tilde{\cal K}^{[ab]}_{\alpha\bar\beta} &=&
\delta_{\alpha\bar\beta} (s+\bar s)^{\nu_{ab}/2-1}
 \prod_{m=1}^3 ( t_m +\bar t_{\bar m} )^{\nu^{(m)}_{ab}-\nu_{ab}/2}
 \prod_{m=1}^3 ( u_m +\bar u_{\bar m} )^{\nu^{(m)}_{ab}-\nu_{ab}/2}  
\eeqn
which is the form that we are going to use in the following. The expression is actually also meant to be diagonal in the 
multiple intersections of the branes $a$ and $b$, i.e. diagonal in the generations of matter multiplets. 
The perturbative type I sectors are mapped 
to perturbative heterotic sectors, whenever $\nu_{ab}=2$, whereas $\nu_{ab}=1$ maps to a non-perturbative 
sector in the same way as the strings among D5-branes in Eq.(\ref{ibanez2}) are. 
In this way, the K\"ahler potential of the heterotic orbifold models is substantially generalized, and 
in particular the distinguished role of the heterotic dilaton negated.   
With this formula, both the D9-D5 and D5-D5 sectors in Eq.(\ref{ibanez2}) can be accomodated. 
Using for instance the shift vector $(1/2,1/2,1)$ reproduces the D9-D5$_3$ potential in Eq.(\ref{ibanez2}) 
whereas $(1/2,1/2,0)$ produces the D5$_1$-D5$_2$ term. In the appendix we have computed the shift vectors for 
a more elaborate example used for modelling a semi-realistic supersymmetric Standard-like Model in 
\cite{Cvetic:2001nr}, which actually 
contains the above D9- and D5-branes as subsectors, showing that they can be consistently implemented 
as special cases within our prescription. \\ 

The effective action derived for heterotic orbifolds, and in particular Eq.(\ref{kaehettw}), is strictly 
only valid in a vicinity of the orbifold point, which is defined as the point in moduli 
space where the locally flat background of the orbifold conformal field theory (CFT) really solves the equations of motion 
of the ten-dimensional string theory. This would require the local cancellation of the RR charges and 
brane tension among the orientifold planes and the D-branes. In general the orientifold planes are located at the fixed loci of 
$\Omega{\cal R}\Theta^k$ (type IIA picture), with $k=0\ {\rm mod}\ N$ referring to the ``pure'' unrotated O6-planes. Via 
$\Omega{\cal R}\Theta^{k}= \Theta^{-k/2}\Omega{\cal R}\Theta^{k/2}$ these fixed loci organize themselves into two orbits 
under the orbifold group $\mathbb{Z}_{2N}$ (or products) \cite{Blumenhagen:2002wn}. The two orbits 
produce two factors in the gauge group of the effective theory, which then will correspond to 
the perturbative and non-perturbative dual heterotic gauge groups, as the D9-branes and D5-branes 
in the simpler $\mathbb{T}^4/\mathbb{Z}_2 \times \mathbb{T}^2$ 
example discussed above. In order to cancel charge and tension 
locally the D6-branes must come to lie on top of these O6-planes in appropriate numbers, 
which unfortunately only leads to non-chiral matter spectra as in \cite{Blumenhagen:1999ev}. But there are more general 
configurations that lead to a cancellation after integrating over the internal space and allow chiral matter. 
These are the ones we are interested in, such that the background that appears in our 
models is strictly speaking not an orbifold and corrections to the effective action may be expected anyway. \\  

It is important to notice that at the orbifold point 
the dependence of the angles on the K\"ahler moduli of the background torus drops out. 
Recall that the (asymmetric) orbifold projection may leave 0, 1 or 3 of the $T_m$ unfixed and that the relative angles in 
principle depend on these via Eq.(\ref{defangle}). In a second step, the supersymmetry condition Eq.(\ref{susy}) will 
fix some or (generically) all of these remaining $T_m$, such that they drop out of the effective action.  
If some of them should instead survive, it is straightforward to substitute a dependence into Eq.(\ref{twkae}) but this  
produces rather awkward expressions when computing the soft parameters such as the scalar masses for this sector. 
More precisely, 
a modulus $T_m$ is not projected out of the spectrum exactly if the orbifold generator $\Theta$ (or generators $\Theta_i$) 
is a reflection along $\mathbb{T}^2_m$, i.e. if $v^{(m)}=1/2$. 
The D-branes on top of the orientifold planes on these tori, that would cancel the charge and tension locally, 
are then given by the degenerate 
cases ${\bf F}_a^{(m)} =0$, pure D9-branes, or ${\bf F}_a^{(m)} = \infty$, localized D5-branes (type IIB picture). 
These two are exactly the choices where the dependence of the angle 
on the modulus $T_m$ drops out, even if they should remain unfixed. 
Thus we see that at the orbifold point no dependence of shift vectors $\nu_{ab}^{(m)}$ on moduli 
exists, which appears very reasonable, since they take the role of modular weights and conformal dimensions. 
Given these considerations, we will actually use Eq.(\ref{twkae}) as a trial metric on heuristic grounds, being aware that 
there may be corrections or modifications in the true K\"ahler metric,\footnote{We would like 
to acknowledge the work of \cite{Lust:2004cx}, where the twisted K\"ahler metric has been computed from first principles 
from string scattering amplitudes, see their section 5.2. 
It appeared some time after the present paper had been published, and 
corrects the above formula (\ref{twkae}), which is only usable for $\frac12$BPS configurations.} 
away from the orbifold point. 
In doing so, we treat the $\nu^{(m)}_{ab}$ as constant free parameters independent of the moduli. 
Together Eq.(\ref{kaepot}) and Eq.(\ref{twkae}) define the full K\"ahler metric of the class of models at hand,
expanded to leading order in the matter fields and subject to the caveats just explained. \\ 


\subsection{The superpotential} 

At the orbifold point the superpotential and D-terms, before considering the deformations by magnetic fluxes 
and the breaking of supersymmtry,  vanish identically. Therefore we are only left with the standard 
superpotential for the fluctuations of the ten-dimensional gauge fields, that induces Yukawa couplings 
but no scalar potential in the four-dimensional theory. 
We start with the classical superpotential known for the heterotic or type I string compactified on a CY 
3-fold ${\cal M}^6$ \cite{Witten:1985xb,Burgess:1985zz,Ferrara:1986qn}  
\be \label{supo} 
W_{\rm tree} = \int_{{\cal M}^6} \Omega_3 \wedge \omega_3 \ , 
\ee
where 
\be 
\Omega_3 = dz_1 \wedge dz_2 \wedge dz_3 \ , \quad 
\omega_3 = {\rm Tr}\ \left( A \wedge dA - \frac{2i}{3} A \wedge A \wedge A \right) 
\ee
are the holomorphic $(3,0)$-form and the Yang-Mills CS-form. Since the gauge field background ${\cal F}_a$ is a 
$(1,1)$ form it does not contribute to Eq.(\ref{supo}) and the only relevant term in $\omega_3$ is 
of the form $f^{ABC} A_i^A A_j^B A_k^C$, where the $A_i^A$ are now the internal components of the 
world volume gauge fields of the D9-branes. Upon noting that 
\beqn \label{supoaa}
\Omega_3 \wedge \omega_3 &=& 2 f^{ABC} ( u_1 A^A_1 +i A^A_2 ) (u_2 A^B_3 + i A_4 )(u_3 A^C_5 + iA^C_6) \, 
   dx_1 \wedge \ \cdots \ \wedge dx_6 \non
&=& 2 f^{ABC} C_{1}^A C_{2}^B C_{3}^C  \, dx_1 \wedge \ \cdots \ \wedge dx_6
\ , 
\eeqn
one finds that the trilinear couplings in $W_{\rm tree}$ are actually independent of the moduli $u_m$, 
once the matter fields have been expressed in terms of the K\"ahler coordinates 
$C^{[aa]}_{m}$ of the first equation of Eq.(\ref{opkae}). Though the expression Eq.(\ref{supoaa}) formally 
looks like the standard commutator form of ${\cal N}=4$ supersymmetry, the orientifold projection on the 
Chan-Paton indices effectively breaks this symmetry \cite{Douglas:1998xa}. 
The surviving couplings are then among the 
components $(\sigma^1\sigma^2 + \sigma^2\sigma^1)\sigma^3$ in the $U(2)$ basis introduced below 
Eq.(\ref{gaugegr}). \\ 

For the open string fields that connect different branes 
the world volume gauge flux shifts the ground state energy of the NS sector such that only one 
of the three complex coordinate fields $C^{[aa]}_{m}$ survives as $C^{[ab]}$, when a stack splits into two. 
Therefore the only terms in Eq.(\ref{supo}) that contain only massless fields, but no massive scalars, are of the form 
\be 
C^{[ab]} C^{[bc]} C^{[ca]} + C^{[ab]} C^{[ba]} C_{m}^{[aa]} \ . 
\ee
A very similar form, again for D9- and D5-brane only, has been deduced 
from open string splitting arguments \cite{Berkooz:1996dw,Ibanez:1998rf}. 
Together, the trilinear couplings in $W_{\rm tree}$ are of the 
form  
\be \label{supoall} 
{\rm Tr} \left(  C^{[aa]}_{1}C^{[aa]}_{2}C^{[aa]}_{3} + C^{[ab]} C^{[bc]} C^{[ca]}
+ C^{[ab]} C^{[ba]} C_{m}^{[aa]} \right) 
\ee
and do not appear to depend on the closed string 
moduli. On the contrary, in \cite{Cremades:2003qj,Cvetic:2003ch,Abel:2003vv}
the mirror symmetric situation with intersecting branes in type IIA orientifolds has 
been evaluated directly by performing a non-perturbative summing over world sheet instanton contributions, 
and in principle a dependence of the Yukawa couplings on the moduli of the dual torus 
arises, which would translate into a dependence on the $u_m$ in our case.  
This appears in fact before putting the matter fields into the proper form Eq.(\ref{opkae}) and it is 
not clear to us if and how the results of \cite{Cremades:2003qj,Cvetic:2003ch,Abel:2003vv} 
are compatible with Eq.(\ref{supoall}) or with \cite{Witten:1985xb,Burgess:1985zz,Ferrara:1986qn}. Since the 
complete absence of moduli denpendence would also cause problems in generating mass hierarchies between 
the quark and lepton generations, we formally keep $Y_{\alpha\beta\gamma}=Y_{\alpha\beta\gamma}(u_m)$ in 
Eq.(\ref{softpara}).  
In particular, among the $Y_{\alpha\beta\gamma}$ only three, which we denote 
$Y_{123}, Y_{abc}$ and $Y_{abm}$, referring to the three terms in Eq.(\ref{supoall}), are non-vanishing. \\ 

This reduction determines the classical, visible part of the total effective superpotential $W_{\rm eff}$. 
Since we do not want to specify the mechanism that finally breaks supersymmetry we do not restrict the 
form of $\hat W$ or the bilinear $\mu$-term in Eq.(\ref{supoeff}). We have argued earlier that we would 
favor a situation with $\mu_{\alpha\beta}=0$ in the tree level superpotential. Such a term may eventually be generated by 
the coefficients $\tilde{\cal Z}_{\alpha\beta}$ in the K\"ahler potential however. 


\subsection{D-terms, Fayet-Iliopolous parameters and axions} 

The DBI action Eq.(\ref{dbi}) also contains terms that only involve the internal components of the gauge fields 
through $(\delta {\cal F}^A)_{ij} = f^{ABC} A^B_i A^C_j$ and the background moduli $\{ s,t_m,u_m\}$. These are then 
part of the scalar potential and, since ${\cal F}$ is a $(1,1)$ form in complex notation, 
originate from D-terms and FI terms. 
Since we focus here completely on the breaking of supersymmetry via $F_I$ taking non-vanishing expectation values, 
we assume that all D-terms vanish in the vacuum. Of course, having D-term breaking in these models sounds like 
an interesting alternative to the present approach.\footnote{This is actually the way supersymmetry gets broken if 
D-branes in the hidden obey different or no calibration condition compared to the visible 
branes \cite{Cremades:2002te,Blumenhagen:2002wn}.} 
The ``D-flatness'' will turn out to be implied by the condition 
Eq.(\ref{susy})
and by restricting the open string scalars $C_\alpha^A \lambda^A$ to take values only in the 
Cartan subalgebra of the gauge group. \\ 

This can actually easily be demonstrated to leading order in the field strength ${\cal F}_a$. Since it involves 
the non-abelian DBI action, which is unknown beyond o$({\cal F}^6)$, and reliably tested so far only up to 
o$({\cal F}^4)$, we cannot easily find the full condition 
Eq.(\ref{susy}). The leading term in $\tilde {\cal F}_a = {\cal F}_a + \delta{\cal F}_a$ 
is simply the YM Lagrangian 
\be 
- \frac{1}{4 g_{10}^2} {\rm Tr} \int d^{10}x\ \sqrt{-g_4} \sqrt{{\rm det} ({\bf n}_a\cdot G)} e^{-\Phi} 
 {\rm tr}  \left( G^{-1} ( {\cal F}_a + \delta {\cal F}_a ) \right)^2 
\ee
because tr$(G^{-1}\tilde{\cal F})=0$ in Eq.(\ref{dettr}). After rescaling to Einstein frame, and 
using $\tilde{\cal F}_{m\bar m} =  (u_m+\bar u_{\bar m}) \tilde{\cal F}^{(m)}_{12}$, one then gets
\beqn \label{kin-dterm}
&& \sum_a N_a \sqrt{\frac{{\rm det} ({\bf n}_a\cdot G)}{{\rm det} (G)^2}} e^{3\Phi} 
{\rm tr}  \left( G^{-1} ( {\cal F}_a + \delta {\cal F}_a ) \right)^2 ~= \\ 
&&
\hspace{2.5cm} 
\frac{16}{s+\bar s} \sum_a N_a \prod_{m=1}^3 {\bf n}_a^{(m)} 
\left( \sum_{m=1}^3 \frac{ ({\cal F}_a)_{m\bar m} + (\delta{\cal F}_a)_{m\bar m} }{
  (t_m+\bar t_{\bar m})(u_m + \bar u_{\bar m})} \right)^2 - 4g_{10}^2 {\cal T}_{\rm O5} \  . 
\nonumber 
\eeqn
The extra term ${\cal T}_{\rm O5}$ is opposite equal to the sum of the tensions of all O5$_m$-planes and arises after applying 
the tadpole constraint Eq.(\ref{rrtad}). It therefore cancels out, as does the leading constant term 
in Eq.(\ref{dettr}) 
with the tension of the O9-planes. 
By going to the Cartan basis one can further put the terms involving 
$(\delta{\cal F}_a)_{m\bar m}$ into the form Tr$|C_m|^2$, $C_m = u_m A_{2m-1} + i A_{2m}$, a sum of 
absolute squares of charged complex scalars weighted by their charges. These $C_m$ are the surviving 
scalars of strings between the two stacks, after turning on the gauge flux to split the stack $a$ into two. 
Since 
\be 
2\Re(f_a) = (s+\bar s) + {\rm o}({\cal F})\ , \quad 
\partial_\alpha {\cal K} =  {\cal K}_\alpha = \frac{\delta_{\alpha\bar \beta} \bar C_{\bar \beta}}{ 
  (t_m+\bar t_{\bar m})(u_m + \bar u_{\bar m})} + {\rm o}({\cal F})\ , 
\ee
this is just of the expected form 
\be 
(D_a + \xi_a)^2 \sim \frac{1}{\Re(f_a)} \left( q^a_\alpha {\cal K}_\alpha C_\alpha + \xi_a \right)^2 \ , 
\ee
to the given leading order in ${\cal F}_a$. 
The $q^a_\alpha$ are the charges of $C_\alpha$ under the $U(1)$ labelled by $a$, and we read off the FI-term 
\be 
\xi_a = \sum_{m=1}^3 \frac{({\cal F}_a)_{m\bar m}}{(t_m+\bar t_{\bar m})(u_m + \bar u_{\bar m})} \ . 
\ee
The supersymmetry condition is 
\be 
\xi_a = \sum_{m=1}^3 \frac{{\bf F}_a^{(m)}}{t_m+\bar t_{\bar m}} = 0 + {\rm o}({\cal F}_a) \ , 
\ee
which is just Eq.(\ref{susyflux}) to leading order in ${\cal F}_a$. This condition is in fact the Donaldson-Uhlenbeck-Yau 
condition $G^{m\bar m} {\cal F}_{m \bar m} = 0$, well known for the heterotic string \cite{Witten:1985bz}. 
The full derivation of Eq.(\ref{susyflux}) from the 
D-term would evidently involve terms of order ${\cal F}^4$ and ${\cal F}^6$. 
In terms of pure Yang-Mills theory this scenario has been studied as early as in \cite{Nielsen:1978rm,Randjbar-Daemi:1982hi}, 
in the context of identifying these 
scalars with Higgs fields in four dimensions. In the YM approximation, some components always get a negative squared 
mass, but in the full string spectrum, this does not have to happen, see e.g. \cite{Rabadan:2001mt}. \\

The pure FI parameter $\xi_a$, 
however, can already be derived from the abelian DBI, and thus Eq.(\ref{susyflux}) can be recovered from an 
equation $\xi_a=0$, and only the fluctuations of charged scalars require the non-abelian action. 
This was also discussed in \cite{Blumenhagen:2003vr} and is based on the exact conditions for $\kappa$-symmetry 
of the abelian DBI action, derived in \cite{Marino:1999af}. 
The relevant constraints on $\tilde {\cal F}_a$ and the K\"ahler form 
\be
J = i \sum_{m=1}^3 \frac{\Re(T_m)}{\Re(U_m)} dz_m \wedge d\bar z_{\bar m} 
 = \sum_{m=1}^3 \Re(T_m) dx_{2m-1} \wedge dx_{2m} 
\ee
are 
\be 
\frac12 J \wedge J \wedge \tilde {\cal F}_a 
 - \frac16 \tilde {\cal F}_a \wedge \tilde {\cal F}_a  \wedge \tilde {\cal F}_a = 0 \ , 
\ee
for all $a$, 
and $\tilde{\cal F}_a$ has to be of type $(1,1)$. For the constant background, setting fluctuations to zero, 
\be 
{\cal F}_a = \sum_{m=1}^3 {\bf F}_a^{(m)} dx_{2m-1} \wedge dx_{2m} \ , 
\ee
and this reproduces Eq.(\ref{susyflux}).  \\

It is also interesting to determine the fate of the imaginary parts of the complex scalar 
moduli fields $s$ and $t_m$, which are components of the RR 2- and 6-form. The Chern-Simons part Eq.(\ref{csaction}) 
of the brane action contains couplings $C_2 \wedge \tilde{\cal F}^4$ and $C_6 \wedge \tilde {\cal F}^2$ 
that reduce to linear axionic couplings of $a_0$ and $a_{m}$, defined above Eq.(\ref{supotree}),  
to $\tilde{\cal F}$ in four dimensions \cite{Cremades:2002te}. Together with the kinetic terms $(da)^2$ 
of the axions, these couplings give St\"uckelberg masses to the respective gauge bosons, i.e. the kinetic 
terms after a duality transformation trading $\{ C_2, C_2^{(m)} \}$ for 
$\{ a_0 , a_m\}$ turn into $(da + A)^2$ and the axion can be gauged away through 
$A \rightarrow A - da$, leaving a mass term for the gauge field $A$. 
Precisely this mechanism was used in \cite{Aldazabal:2002py} 
to propose a solution for the strong CP problem by promoting the 
QCD $\theta$-parameter to a dynamical axion field that has additional St\"uckelberg couplings and can be gauged away then. 
Therefore, the axions as well decouple as longitudinal components of massive vectors 
from the effective action, together with their moduli partners, when 
gauge fluxes are turned on. An important point noticed in \cite{Ibanez:2001nd} is that the abelian gauge bosons may even  
decouple if the gauge symmetry is not anomalous. For an anomalous $U(1)_a$ the Green-Schwarz mechanism actually 
requires a second axionic coupling of the $a$ to $\tilde{\cal F}_a^2$, which then allows the cancellation of triangle 
anomalies by axionic contributions.  
To see that the mechanism really works out properly, by which we mean, moduli and axions decouple in a one to one fashion, 
one has to verify that the axionic 
couplings come exactly in the same patterns that the D-terms arise for the $\Re(T_m)$. We define coupling 
constants \cite{Cremades:2002te}
\beqn 
c_{a}^{(0)} C_2 \wedge \delta{\cal F}_a 
 &=& C_2 \wedge \delta{\cal F}_a \int_{{\cal W}_a} \frac{1}{6} {\cal F}_a \wedge {\cal F}_a \wedge {\cal F}_a \ , \non
c_{a}^{(m)} C_2^{(m)} \wedge \delta{\cal F}_a &=& C_2^{(m)} \wedge \delta{\cal F}_a \int_{{\cal W}_a} C_4^{(m)} 
  \wedge {\cal F}_a^{(m)} d\xi_{2m-1} \wedge d\xi_{2m} \ . 
\eeqn
In other words, $c_{a}^{(0)}$ is the coupling of the universal axion, the partner of $s+\bar s$ and 
$c_{a}^{(m)}$ the coupling of the partner of $t_m +\bar t_{\bar m}$. 
Since the number of stacks will be at least four or larger, the generic situation seems to be that all four axions 
have non-vanishing couplings to one linear combination of $U(1)_a$ gauge bosons. However, the calibration 
condition Eq.(\ref{susyflux}) implies that one can find a linear combinations of the four axions that 
decouples from the gauge fields. Setting the internal components of $C_6$ to 1, and defining 
$C_2' = (s+\bar s)C_2,\ C_2^{(m)'} = -(t_m +\bar t_{\bar m}) C_2^{(m)}$, this ``diagonal'' axion
is given by the direction $C'_2 = C_2^{(1)'}= C_2^{(2)'}= C_2^{(3)'}$, since then its couplings vanish,  
\be \label{diagax} 
\left( \frac{c_{a}^{(0)}}{\prod_{m=1}^3 \Re(T_m)} C_2' -  
  \sum_{m=1}^3 \frac{c_{a}^{(m)}}{\Re(T_m)} C_2^{(m)'} \right) \wedge \delta{\cal F}_a \Big|_{C'_2=C_2^{(m)'}}  = 0 \ , 
\ee
by Eq.(\ref{susyflux}) and for any $a$. Given the D-flatness, we thus have only three axions participating and the diagonal axion in 
Eq.(\ref{diagax}) survives. The linear 
combination of axions that decouples corresponds to the surviving modulus, due 
to rewriting Eq.(\ref{susyflux}) as 
\be \label{dflat}
\sum_{m=1}^3 \frac{s+\bar s}{t_m+\bar t_{\bar m}} {\bf F}_a^{(m)} 
= \prod_{m=1}^3 {\bf F}_a^{(m)} \ . 
\ee
A variation along the complex direction $s= t_1=t_2=t_3$ leaves the D-flatness Eq.(\ref{dflat}) intact 
as well as Eq.(\ref{diagax}), identifying $\Im(s)=C_2$ and $\Im(t_m)=C_2^{(m)}$. 
It is evident that less generic cases, in which the D-flatness Eq.(\ref{susyflux}) leaves more 
moduli massless, one can also construct more linear combination of axions that survive together 
with the abelian gauge bosons. \\

The above is interesting for several reasons. First it implies that starting from a generic four 
stack model, we precisely expect one massless abelian gauge factor to survive, which is then the unique 
candidate for the hypercharge $U(1)_Y$ of the supersymmetric Standard Model. The other three 
$U(1)$ gauge symmetries decouple from the massless theory and survive as global symmetries only. It also  
follows that three among the seven phases we started with in parametrizing the values of the auxiliary 
fields $F^I$ for $\{s,t_m,u_m\}$, are frozen. 
Technically, this unfortunately does not appear to be too helpful, since it does not imply, that 
any of the phases of $\{s,t_m\}$ actually vanish, but instead only that they are parametrized through a single 
phase along Eq.(\ref{diagax}), but all four being non-vanishing. Only in the situation when all phases are aligned, 
the soft-breaking parameters and the patterns for CP violation will simplify considerably. This situation may arise if the 
scale of supersymmetry breaking is well below the scale where the relative moduli among $s,t_m$ get fixed. Then 
the four complex scalars and auxiliary fields would effectively already be aligned and the phases of the respective auxiliary fields 
in the chiral multiplets would be equal. \\ 


\section{Soft supersymmetry breaking}

We now straightforwardly apply the formulas for computing the soft parameters using the 
expressions given in the previous section. 
First we address the vacuum energy, which we have to assume to take a very small or even vanishing value, 
of course. Note that this in $\mathbb{Z}_2\times \mathbb{Z}_2$ orbifolds this is actually satisfied for any 
model, where the moduli part of the superpotential depends only on either complex structure moduli $u_m$ or 
K\"ahler moduli $t_m$. Then  
the vacuum energy is non-negative, $V_0 \ge 0$, since the potential is of the classical 
no-scale type \cite{Cremmer:1983bf,Ellis:1983sf}. 
In the first case, which would be realized, as long as the effects that drive the breaking are perturbative 
in the string coupling, the contribution of the $t_m$ moduli to the potential cancel the negative term through 
\be \label{cancel} 
\sum_{m=1}^3 \hat {\cal K}^{t_m \bar t_{\bar m}} 
  D_{t_m} \hat W(u_m) \bar D_{\bar t_{\bar m}} \bar{\hat W}(u_m) = 3 |\hat W|^2 \ . 
\ee
The (flat Minkowski) vacuum is then supersymmetric whenever all other contributions to the potential exactly vanish. 
In models with less than three K\"ahler moduli, the situation would possibly look different. 
The case of gaugino condensation in a hidden sector, as considered in \cite{Cvetic:2003yd}, is of 
the opposite type. In the absence of perturbative contributions, 
the effective superpotential is depending on the moduli only through 
the gauge kinetic functions $f_a(s,t_m)$. In that case, the contribution of the generic three $u_m$ moduli can cancel 
the negative term, just as in Eq.(\ref{cancel}) upon substituting $t_m \leftrightarrow u_m$. \\

The models we are considering have at most seven complex scalar moduli fields above the scale where the 
constraint Eq.(\ref{susy}) fixes up to three among $\{s,t_m\}$, so that at least one combination of these 
and three $u_m$ survive. At low energies there will then be at least four free real parameters and phases 
for the values of the respective auxiliary fields, determining the patterns of supersymmetry breaking. 
In the formulas we now always use the maximal number of seven moduli. 
To proceed further we parametrize the auxiliary fields in a standard fashion (see e.g. \cite{Brignole:1997dp}), 
\beqn 
F^s &=& c (s+\bar s) \sin(\theta) e^{-i\gamma_s}\ , \\ 
F^{t_m} &=& c ( t_m + \bar t_{\bar m} ) \cos(\theta) {\bf \Theta}_{t_m} e^{-i\gamma_{t_m}} \ , \quad 
F^{u_m} ~=~ c ( u_m + \bar u_{\bar m} ) \cos(\theta) {\bf \Theta}_{u_m} e^{-i\gamma_{u_m}} 
\nonumber 
\eeqn
with 
\beqn 
c ~=~ C \sqrt{3} M_{3/2} \ , \quad 
C^2 ~=~ 1 + \frac{V_0}{3M^2_{3/2}} \ , \quad 
\sum_{m=1}^3 \left( |{\bf \Theta}_{t_m}|^2 + |{\bf \Theta}_{u_m}|^2 \right) ~=~ 1 \ . 
\eeqn
We further define \cite{Nath:2002nb}
\beqn 
D &=& -\ln(s+\bar s)\ , \quad\quad\quad  e^{-i\rho} ~=~ \frac{ \langle \hat W \rangle }{ |\langle \hat W \rangle|} \ , \non 
f &=& \prod_{m=1}^3 \left( t_m + \bar t_{\bar m} \right) \prod_{m=1}^3 \left( u_m + \bar u_{\bar m} \right) \ . 
\eeqn
%

\subsection{The soft breaking parameters} 

For the gaugino masses Eq.(\ref{gaugmass}) we now obtain 
\beqn 
M_a &=& \frac{c}{2\Re (f_a)} \Big ( \sin(\theta)e^{-i\gamma_s} (s+ \bar s) \prod_{m=1}^3 {\bf n}^{(m)}_a \\ 
&& \hspace{2cm}  
- \frac12 \sum_{m,n,p=1}^3  \gamma_{mnp} \cos(\theta) {\bf \Theta}_{t_m} e^{-i\gamma_{t_m}} ( t_m + \bar t_{\bar m} ) 
  {\bf n}^{(m)}_a {\bf m}^{(n)}_a {\bf m}^{(p)}_a  \Big) \ .  
\nonumber 
\eeqn 
Whenever just a single term in the bracket survives, i.e. for D9- or D5-branes, the dependence of $M_a$ on 
the moduli drops out, but in all other cases the gaugino masses will depend on (the real parts of) $s$ and the K\"ahler 
parameters $t_m$. 
For the scalar mass parameters of extra matter fields in symmetric and anti-symmetric representations we get  
\beqn 
M^{[aa]\, 2}_{m\bar m} &=& M^2_{3/2} + V_0 - F^{I} \bar F^{\bar J} \partial_{I} \partial_{\bar J} 
  \ln \left( \tilde{\cal K}_{m\bar m}^{[aa]} \right) \non 
&=& M^2_{3/2} + V_0 + c^2 \cos^2(\theta) |\Theta_{u_m}|^2 + \Gamma^{[aa]}_{m\bar m}  \ , 
\eeqn
with 
\beqn \label{massmm}
\Gamma^{[aa]}_{m\bar m} &=& 4 |M_a|^2 - \frac{c^2}{(1+\Delta^{(m)}_a)^2} \times \\
&& \hspace{-2cm} 
\times \Bigg( |\sin{(\theta)}e^{-i\gamma_s}
+\cos{(\theta)} \Theta_{t_m}e^{-i\gamma_{t_m}} 
+\frac{1}{2}\cos{(\theta)}\sum_{n,p} \Delta_a^{(m|np)} (\Theta_{t_n}e^{-i\gamma_{t_n}} +\Theta_{t_p}e^{-i\gamma_{t_p}})|^2
\non
&& \hspace{-1cm} 
- 2 (1+\Delta_a^{(m)})\Big( \sin{(\theta)}\cos(\theta) \Theta_{t_m}
\cos(\gamma_s-\gamma_{t_m}) 
\non
&& \hspace{2cm} 
+ \cos^2{(\theta)} \sum_{n,p}\Delta_a^{(m|np)}  \Theta_{t_n}\Theta_{t_p}
\cos(\gamma_{t_n}-\gamma_{t_p}) \Big) \Bigg) \nonumber \ . 
\eeqn
We note that one hallmark of the case of general background world volume flux 
is the appearance of cross terms with $\Theta_{t_n}\Theta_{t_p}$, 
as they are absent for only D9- and D5-branes. 
For completeness, we state the trilinear coupling parameters for the $C_{m}^{[aa]}$ fields,  
\beqn \label{tri} 
A^0_{123} - \bar c F^I \partial_I \ln( Y_{123} ) &=& \\ 
&& \hspace{-3cm} 
- c \frac{e^{-i\rho +\frac{D}{2}}}{\sqrt{f}}
\Bigg( 
\sum_{m=1}^{3} \frac{\Delta_a^{(m)}}{1+\Delta_a^{(m)}}
( \sin{(\theta)} e^{-i\gamma_s} +\cos{(\theta)} \Theta_{t_m}e^{-i\gamma_{t_m}})
\non 
&& \hspace{-3cm} 
+ \frac{6}{c}M_a - 2 \sin{(\theta)} e^{-i\gamma_s} 
- \cos{(\theta)} \sum_{m,n,p=1}^3 \frac{\Delta_a^{(m|np)}}{1+\Delta_a^{(m)}}
(\Theta_{t_n}e^{-i\gamma_{t_n}} + \Theta_{t_p}e^{-i\gamma_{t_p}}) \Bigg) \nonumber \ .
\eeqn
The most interesting matter sector is the one which involves the bifundamentals, the squarks and sleptons.
Here for $M^{[ab]2}$ we find 
\beqn \label{bi-mass}
M^{[ab]2} &=& M_{3/2}^2+V_0 - c^2\sin^2{(\theta)}\Big( 1-\frac{\nu_{ab}}{2} \Big)
\non 
&&
- c^2 \cos^2{(\theta)} \sum_{m=1}^{3} \nu_{ab}^{(m)}
\Big( \frac{1}{2}- |\Theta_{t_m}|^2 -
 |\Theta_{u_m}|^2 \Big) 
\eeqn
>From this expression one can derive simple sum-rules. In a rather isotropic case, when all 
$\nu_{ab}^{(m)} \sim \nu_{ab}/3$ one finds e.g. 
\beqn 
M^{[ab]2} &\sim& M_{3/2}^2+V_0 - c^2 \left( 1- \frac{\nu_{ab}}{2} - \cos^2{(\theta)} \left( 1+ \frac{\nu_{ab}}{3} \right) \right) \ .
\eeqn
While the general expression for the scalar masses is already independent of the phases of the $F^I$, this 
formula does only involve the single angle parameter $\theta$ which distinguishes between the popular scenarios of 
dilaton or moduli domination. 
For the trilinear couplings involving three bifundamentals we
get
\beqn \label{bi-tri1}
A_{abc}^0 - \bar c F^I\partial_I \ln( Y_{abc}) &=& c \frac{e^{-i\rho +\frac{D}{2}}}{\sqrt{f}}
\left( \Big( 2-\frac{\nu_{ab}+ \nu_{bc} + \nu_{ca}}{2} \Big) \sin{(\theta)} e^{- i\gamma_s}
\right. \\ 
&& \left. \hspace{-5.5cm} 
- \cos(\theta) \sum_{m=1}^{3} 
\left( 1+\nu_{ab}^{(m)}+ \nu_{bc}^{(m)} +\nu_{ca}^{(m)}
-\frac{\nu_{ab}+ \nu_{bc} + \nu_{ca}}{2} \right)
(\Theta_{t_m}e^{-i\gamma_{t_m}}
+\Theta_{u_m} e^{-i\gamma_{u_m}})
\right) \nonumber \ . 
\eeqn
When again the $\nu_{ab}^{(m)}$ etc. have all equal elements $\nu_{ab}/3=\nu_{bc}/3=\nu_{ca}/3$
 the relation of Eq.(\ref{bi-tri1}) simplifies so that  
\beqn \label{bi-tri2}
A^0_{abc} - \bar c F^I\partial_I \ln( Y_{abc}) &\sim& c \frac{e^{-i\rho +\frac{D}{2}}}{\sqrt{f}}
\Bigg( \left( 2-\frac{3\nu_{ab}}{2}\right) \sin{(\theta)} e^{- i\gamma_s}
\\
&& 
-\cos(\theta) \left( 1-\frac{\nu_{ab}}{2} \right)
\sum_{m=1}^{3} \left( \Theta_{t_m}e^{-i\gamma_{t_m}}
+\Theta_{u_m}e^{-i\gamma_{u_m}}\right) 
\Bigg)
\nonumber 
\eeqn
We also record here the trilinear couplings involving only two bifundaments,
\beqn \label{bi-tri3}
A_{abm}^0 - \bar c F^I\partial_I \ln( Y_{abc}) &=& 
c \frac{e^{-i\rho +\frac{D}{2}}}{\sqrt{f}}
\Bigg( 
-\frac{1}{c} M_a + \sin{\theta} e^{-i\gamma_s} (2-\nu_{ab}) 
\\
&&  \hspace{-5.5cm} 
+ \cos{(\theta)} \Big( 
\Theta_{t_m}e^{-i\gamma_{t_m}}+\Theta_{u_m}e^{-i\gamma_{u_m}}
+\sum_{n=1}^{3}(-1-2\nu_{ab}^{(n)}+\nu_{ab})
(\Theta_{t_n}e^{-i\gamma_{t_n}}
+\Theta_{u_n}e^{-i\gamma_{u_n}})\Big)
\non
&& \hspace{0cm} 
-\frac{\Delta_a^{(m)}}{1+\Delta_a^{(m)}}
(\sin{(\theta)}e^{-i\gamma_s} +\cos{(\theta)} \Theta_{t_m}e^{-i\gamma_{t_m}})
\non
&& \hspace{0cm} 
+ \frac{\cos(\theta)}{2} \sum_{n,p}\frac{\Delta_a^{(m|np)}}{1+\Delta_a^{(m)}}
(\Theta_{t_m}e^{-i\gamma_{t_m}}+\Theta_{u_m} e^{-i\gamma_{u_m}}) \Bigg)
\nonumber 
\eeqn 
Finally, we compute the parameter relevant for the Higgs bilinear term 
$B^0_{ab} = B^0_{[ab][ba]}$, which refers to a sector of open strings between two given branes $[ab]$. 
\beqn \label{bi-bterm}
B^0_{ab} &=& 
c \frac{e^{-i\rho+\frac{D}{2}}}{\sqrt{f}}
\Bigg( 
\sin{(\theta)}e^{-i\gamma_s}\left( 1-\nu_{ab} +
(s+\bar s)\partial_s \ln(\mu_{ab}) \right)
\\
&&
- \cos{(\theta)}\sum_{m=1}^{3}\Theta_{t_m} e^{-i\gamma_{t_m}}
\left(1 + 2\nu_{ab}^{(m)}-\nu_{ab} 
-(t_m+\bar t_{\bar m})\partial_{t_m}\ln(\mu_{ab}) \right)
\non
&&
- \cos{(\theta)}\sum_{m=1}^{3}\Theta_{u_m} e^{-i\gamma_{u_m}}
\left( 1+2\nu_{ab}^{(m)}-\nu_{ab}
-(u_m+\bar u_{\bar m})\partial_{u_m}\ln(\mu_{ab}) \right)
\Bigg) \ . 
\nonumber
\eeqn
When $\mu_{ab}$ is independent of the moduli, as we have argued above may be natural, 
the terms proportional to derivatives drop out. An application of these
general formulae to a specific model is given in appendix A. 


\subsection{Simplifications for $s=t_1=t_2=t_3$} 

Since it appears possible that the supersymmetry constraints Eq.(\ref{susy}) fix the four mentioned 
complex moduli fields to be effectively aligned, and also the auxiliary fields in their respective chiral multiplets, 
we briefly investigate the simplifications that arise when replacing the four fields by a single one. Of course, 
we do not mean that the vacuum expectation values of $s$ and $t_m$ are equal, but just that the surviving 
modulus points into the diagonal direction, which we call $\lambda$. 
This is perhaps a very naive way to implement the integrating out of 
the relative coordinates among the $s,t_m$ by just substituting $\lambda$, 
but the analysis here may give a rough idea of what can happen in this situation. The auxiliary field is 
$F^\lambda= c(\lambda + \bar \lambda) \Theta_\lambda e^{-i\gamma_\lambda}$. 
One can then observe that 
\beqn
\Delta^{(m)}_a \propto ({\bf F}_a^{(m)})^2 \ , 
\eeqn
independent of the remaining moduli. It also follows immediately that the gaugino masses are completely determined by 
the value of $F^\lambda$, 
\be
M_a = \frac{1}{2\Re(f_a)} F^\lambda \partial_\lambda f_a = c \Theta_\lambda e^{-i\gamma_\lambda} 
\ee
and all explicit dependence on the world volume gauge fields drops out. This follows already from 
$f_a=f_a(\lambda)$ only depending linearly and holomorphically on $\lambda$,
 but no $u_m$.  
Thus the gaugino masses are approximately universal. The simplifications for the other parameters can in fact also 
be quite dramatic. The scalar masses in the $[aa]$ sectors simplify so that  
\beqn 
M_{m\bar m}^{[aa]\, 2} = M_{3/2}^2 + V_0 + c^2 \cos^2(\theta) |\Theta_{u_m}|^2 + 2 c^2 |\Theta_\lambda |^2 \ , 
\eeqn
still denpending on the $F^{u_m}$, but universal in $a$, and therefore independent of ${\cal F}_a$. 
An interesting point is that the cross terms in Eq.(\ref{massmm}) 
drop out in this simpler case. In the same way their trilinear couplings $A^0_{123}$ become independent of the stack 
of branes, and one would predict a set of extra matter at some universal mass in anti-symmetric and symmetric representations 
of the gauge group for this simple scenario. 
It is very easy to get the specialized formulas for the squarks and sleptons as well. These, fortunately, 
do not lose their dependence on the type $[ab]$ of intersection, where the fields are localized, which enters 
through the $\nu_{ab}^{(m)}$, and they still depend 
on $F^{u_m}$, such that no qualitative reduction of the complexity of possible solutions takes place. The number 
of independent parameters is of course reduced by three. 


\section{Gauge unification, FCNC, EDM, Dark matter} 

In this section we discuss a number of phenomena within intersecting brane
models which are of interest in model building in particle theory.
Specifically we discuss issues regarding gauge coupling unification,
flavor changing neutral currents, CP violation, the electric dipole
moments, and dark matter. Another interesting application of the
 intersecting brane model concerns proton decay via dimension
 six operators \cite{Klebanov:2003my}. However, this topic will not
 be discussed here. A more systematic analysis and a deeper investigation 
 of these topics is left to a future work. 

\subsection{Gauge coupling unification} 

The unification of gauge couplings using standard renormalization group running is one of the facts 
about the minimal extension of the Standard Model, that is usually considered among its most attractive 
features. Starting from LEP data and using the particle spectrum of the MSSM the couplings meet at 
$M_{\rm GUT} = 2 \times 10^{16}$GeV. 
A D-brane model derived 
from string theory, that tries to construct the MSSM already at a high scale, without any extra matter or gauge 
group, will have to be able to reproduce this unification pattern, since 
otherwise it would lead to the wrong couplings at low energies, just turning the evolution of 
couplings around. So even if we do not expect a unification of the gauge group, e.g. an enhancement towards 
$SU(5)$ or $SO(10)$ at the GUT scale, since we are still dealing with separate stacks of D-branes then, we 
have to worry about unification as an accidental property of our models. This argument does, however, not 
apply in practice so far, since all examples of models known involve extra sectors with additional matter 
and gauge factors. Their dynamics at intermediate scales would substantially affect the running of the couplings, and 
thus the apparant unification of couplings may be an illusion and spoiled by these effects. \\ 

In any case, we now investigate the minimal requirements that would arise from imposing 
gauge unification (see \cite{Blumenhagen:2003jy,Blumenhagen:2003qd,Chamoun:2003pf}). 
What is meant as an attempt to further constrain the models and increase their predictive power will turn out 
to lead to a system of overconstraining conditions in a sufficiently generic case. 
Using $SU(5)$ conventions, the 
unification actually reads 
\be \label{gcplequ}
\frac{1}{g_3^2(M_{\rm GUT})} = \frac{1}{g_2^2(M_{\rm GUT})} = \frac53 \frac{1}{g_Y^2(M_{\rm GUT})} \ , 
\ee
where the $g_i$ denote the couplings of $SU(3)$, $SU(2)$ and $U(1)_Y$. The latter will actually be some 
linear combinations $\sum_a c_a U(1)_a$ 
of the $U(1)_a$ that are present in the total model, one per stack of branes. 
The starting point of \cite{Blumenhagen:2003jy} is that Eq.(\ref{gcplequ}) is compatible with the 
condition that arises by embedding the hypercharge $U(1)$ into the original abelian gauge 
symmetries that live on the $U(3)\times U(2)\times U(1)^n$ stacks to start with. \\ 

One may now distinguish three cases. $i)$ This is the apparantly most generic case. 
There is at least one stack of branes such that all three ${\bf F}_a^{(m)}$ are 
non-vanishing, which does not even have to be a stack in the visible sector of the model. Then Eq.(\ref{susyflux}) 
will fix the overall volume of the internal space and, if not very special accidents appear, also all ratios 
$\Re(T_1):\Re(T_2):\Re(T_3)$, i.e. fix $\{s+\bar s, t_m+\bar t_{\bar m}\}$ except for a simultaneous rescaling 
of all four. Now Eq.(\ref{gcplequ}) means 
\be
\Re(f_3(s,t_m)) = \Re(f_2(s,t_m)) = \frac53 \sum_a c_a \Re(f_a(s,t_m)) \ . 
\ee
Since the $f_a(s,t_m)$ are linear functions in $\{s,t_m\}$ an overall rescaling does not affect this relation. 
Therefore, the D-flatness Eq.(\ref{susyflux}) already fixes all freedom in the relevant parameters that could be 
used to unify the gauge couplings, and Eq.(\ref{gcplequ}) would be very accidental. $ii)$ All stacks have at least 
one $m$ for which ${\bf F}_a^{(m)}=0$. Though this sounds less generic, it is actually the case in all the 
examples to be found in the literature. The reason is probably just 
that the practical search for models is very much simplified, if 
Eq.(\ref{susyflux}) can be turned into a linear relation. But we do not see any reason why this situation should be 
favored on general grounds. Now $s$ drops out of Eq.(\ref{susyflux}) and the overall volume remains undetermined, 
while the ratios $\Re(T_1):\Re(T_2):\Re(T_3)$ are fixed as long as no further relations among the conditions exist. 
This situation provides one single free parameter in Eq.(\ref{gcplequ}), not enough to satisfy two relations. Unification 
would again be unnatural, and is not achieved in the cases considered in \cite{Cvetic:2001nr}. 
$iii)$ If in case $ii)$ there are further relations among the conditions Eq.(\ref{susyflux}) 
then two parameters may survive, and only one ratio of $\Re(T_m)$ is fixed. 
This is actually the case considered in \cite{Blumenhagen:2003jy}, and 
applies to some of the simpler examples given in the literature. Imposing the extra constraint Eq.(\ref{gcplequ}) is then 
just enough to fix the remaining two K\"ahler parameters. Thus, only in the most symmetric case, unification appears 
possible generically. \\ 

If a class of examples of the type $i)$ were found that produced precisely the MSSM spectrum, 
imposing the extra constaint Ref.(\ref{gcplequ}) could then maybe point toward the ``true'' solution. Otherwise, 
one should look for more compelling reasons to consider special examples where some flux quantum numbers 
vanish and simplifying symmetries exist. As a final remark, note that the difficulty in obtaining a unification 
of gauge couplings in the present class of D-brane models is about the opposite of the situation in the 
heterotic string, where grand unification is somehow automatic.

\subsection{Flavor  changing neutral currents}

In a general situation of soft supersymmetry breaking 
the parameters in the soft breaking terms of the effective Lagrangian can be arbitrary for 
different generations of matter multiplets. This kind of anisotropy leads to phenomenological problems, since then 
interferences of these fields no longer cancel out. A prominent example of such effects are flavor  changing neutral 
currents that for instance lead to unacceptable rates for transitions which contribute to the same process as the 
$\Delta S=2$ box diagram that allows for CP violation in the K-system 
(see \cite{Ellis:1981ts,Donoghue:1983mx,Nilles:1983ge}\footnote{A completely different source for FCNC was 
discussed in \cite{Abel:2003fk} in the context of brane models with supersymmetry breaking at a low string scale, where the 
massive fields can contribute as well.}). 
The absence of such transitions puts 
bounds on the differences of the masses of squark and slepton doublets, but in principle not the 
anti-squark singlets. 
In the present setting of brane models however, all three generations are generically on a very symmetric footing. 
They arise from multiple intersections of the same two stacks of branes, and therefore the relevant K\"ahler metrics 
are equal. The only difference could arise in their Yukawa couplings $Y_{\alpha\beta\gamma}$, which would only lead to 
mass differences comparable to those of the quarks and leptons of the Standard Model itself. 
The only exception to this is indeed the interesting case of the quark doublets, where in a large variety of models 
the three generations split into two from one set of intersections and a third from an extra intersection with potentially 
different relative angles and K\"ahler metric. The reason for this to happen is the fact that this allows to circumvent 
a problem related to the anomaly cancellation within the $U(2)$, which demands the number of positively and negatively 
charged doublets to be equal. This can be arranged if the two $({\bf 3},{\bf 2})$  carry charge plus, 
the one extra $({\bf 3},{\bf 2})$ 
charge minus, and the three lepton doublets
 $({\bf 1},{\bf 2})$ also minus \cite{Ibanez:2001nd}. \\
 
For the FCNC constraints to be satisfied for the first two generations, the 
squarks have to be essentially degenerate with mass differences of the order 
of the charm quark mass. The constraints  on the third generation 
squark masses on the Higgs masses consistent with FCNC constraints
are far less stringent. Thus if we label the first two generation
squark masses by $M^{[ab]}_q$ and the third generation squark masses by
 $M^{[cd]}_q$ then the condition $M^{[cd]2}_q$ = $(1+\delta)M^{[ab]2}_q$  
is consistent with FCNC constraints with $|\delta|\leq 1$ \cite{Matalliotakis:1994ft,Nath:1997qm}.
The above translates to the following conditions on the intersection angles
\beqn\label{fcnc} 
&& \hspace{-1cm}
\delta \left( 
 \frac13 +\frac12 \sin^2(\theta) ( \nu_{cd} -2 ) - \cos^2(\theta) \sum_{m=1}^3 \nu_{cd}^{(m)} 
  \left( \frac12 - |\Theta_{t_m}|^2 - |\Theta_{u_m}|^2 \right) \right) = \\
&& \frac12 \sin^2(\theta) ( \nu_{cd} - \nu_{ab} ) + \cos^2(\theta) \sum_{m=1}^3 ( \nu_{cd}^{(m)} - \nu_{ab}^{(m)} )  
  \left( \frac12 - |\Theta_{t_m}|^2 - |\Theta_{u_m}|^2 \right) \ , \nonumber 
\eeqn
where we have set $V_0=0$ in writing the above constraint.
Now Eq.(\ref{fcnc}) imposes only mild constraints on the intersection angles
for the third generation compared to the the first generation. Depending on the breaking scenario, 
they can become sharpened. As an example, complete dilaton domination with $\cos(\theta)=0$ would 
put rather stringent bounds on $\nu_{ab}$ and $\nu_{cd}$. One also has 
similar FCNC constraints for the Higgs doublet masses at the
string scale. Thus replacing $M^{[cd]}_{q}$ with $M^{[ef]}_{H}$,
where $M^{[ef]}_{H}$ is the Higgs mass for the generations localized at 
the intersection of $[ef]$, then Eq.(\ref{fcnc}) holds again with
$\delta$ replaced by $\delta_H$, where FCNC constraints are consistent
 with $|\delta_H|\leq 1$. Again the constraints on
 the intersection angles in this case are rather mild. The anisotropy between the first two and the third generation 
could only become dangerous if experimental bounds were sharpened, comparable to the first two generations, 
which are completely degenerate in the brane models at hand.


\subsection{CP-violation} 

The soft breaking sector of the intersecting brane models contains 
seven phases, i.e. $\gamma_s$, $\gamma_{t_m}$, $\gamma_{u_m}$ 
$(m=1,2,3)$.  The typical size of these phases is o$(1)$ and they
produce large  effects on the electric dipole
moments (EDMs) of the electron and of the neutron 
(complete expressions for these can be found in 
\cite{Ibrahim:1997nc,Ibrahim:1998je}).  Thus the moduli
 phase generated EDMs may in general exceed the very sensitive 
experimental bounds 
on the electric dipole moments of the electron, the
neutron and of the Hg-atom \cite{Commins:gv,Harris:jx,Lamoreaux:1986xz}.
 These models can be made 
compatible with experiment either via mass suppression \cite{Nath:dn}
or via the cancellation mechanism \cite{Ibrahim:1999af,Brhlik:1999ub}. 
The phases can affect a large number of phenomena accessible to
experiment (for a recent review see \cite{Ibrahim:2002mu}).
These include  sparticle decays, $g_{\mu}-2$, proton decay, 
$ B_{(s,d)}^0\rightarrow l+ l-$,and baryogenesis \cite{Carena:2002ss}, to name a few.  There is one case in which
drastic simplication occurs on the dependence on the phases. 
If all phases are equal,  
\beqn
\gamma_s=\gamma_{t_1}=\gamma_{t_2}=\gamma_{t_3} \equiv \gamma
\eeqn
one finds that the common phase factors out of 
the gaugino masses
and in view of the discussion of sec 4.2 one may write
\beqn 
M_a &=& M_{{1}/{2}} e^{-i\gamma} \ . 
\eeqn 
Here the gaugino mases are universal independent of the brane
stack. Further, the common phase can be rotated away from the
gaugino sector by redefinition of fields although it will appear in
other sectors of the theory. Thus  the only phases that
 remain in this case are $\gamma$ and $\gamma_{u_m}$.
An important phase that enters in physical processes is the
$\mu$-phase $\theta_{\mu}$, the phase of the term $\mu_{\alpha\beta}$ in the superpotential Eq.(\ref{supoeff}). 
In principle it is determined in terms of 
the fundamental moduli phases, but we have not specified the $\mu$-term here, 
in case it is non-vanishing.If one considers the subset of intersecting brane models where
  $\theta_{\mu}$ vanishes or is very small (i.e. of order $10^{-2}$, which may be natural, since we have 
argued that it may be vanishing to leading order), 
then the only avenue for the phases to enter the physical processes is via the trilinear parameter.
  Further,  an arrangement that the trilinear parameters for the
  first two generations are relatively real while the 
   third generation $A^0$ is complex will satisfy the experimental 
   EDM limits  for a significantly large $A^0$ phase\cite{Chang:1998uc}.
    This scenario
has some interesting features. Thus, for example, there will
be no CP-phase dependent contribution from the dominant 
chargino-sneutrino contribution to $g_{\mu}-2$, 
because the chargino mass is independent of the phases
in this  case and so is the sneutrino mass. However, the phase of  
the third generation $A^0$ can still produce large effects. Such effects
will be visible in the decay of the stops once they are produced at 
the Fermilab  Tevatron and at the Large Hadron Collider at CERN.
%


\subsection{Dark matter} 

The recent data from WMAP \cite{bennett} indicates that there is a significant
cold dark matter component to the dark matter- dark
energy in the universe. Like 
SUGRA models, the intersecting brane models with
R-parity constraint have the possibility that the 
lowest mass  neutralino could be the lowest mass 
supersymmetric particle  (LSP) and a candidate for cold dark matter. 
Recent analyses of the WMAP data indicate
that the accurate WMAP results produce a strong correlation 
between the sfermion and gaugino masses and the
allowed mass range can extend to even 
tens of TeV \cite{Chattopadhyay:2003xi} (for a review 
see \cite{Lahanas:2003bh}).
Previous analyses indicate  that  brane models 
can indeed allow for the desired amount of dark
matter \cite{Corsetti:2000yq}. Since the pattern of 
soft breaking in the intersecting brane models is 
essentially a modification of that for the parallel
brane case, one expects that the intersecting 
brane model will sustain dark matter in sufficient
amounts to be compatible with the WMAP data. As in 
the case of heterotic string \cite{Nath:2002nb}
the constraints of radiative breaking of the electroweak symmetry
in the intersecting
brane model will also determine $\tan(\beta)$.  In this case the
 analysis of dark matter would be much more predictive 
 compared to the SUGRA models.  It would
in fact be interesting to investigate this possibility
in further detail for the case of intersecting brane models. 


\section{Conclusions and prospects}
 
In this work we have investigated the  effective action and
soft breaking in a generic class of intersecting brane models with
${\cal N}=1$ supersymmetry. There are in general two equivalent approaches 
to such contructions, i.e. models based on intersecting D6-branes 
in type IIA  or alternatively to use constructions with
D9-branes with magnetic flux on their world volume in type IIB. 
 The type IIB framework is found more convenient to construct the 
 K\"ahler metric, the gauge kinetic function and the superpotential, 
 and we have followed  this approach in the present
 work. Thus one of the main results of this paper are Eq.(\ref{kaepot})
and Eq.(\ref{twkae}) which give the K\"ahler metric for chiral matter 
in the intersecting brane case and reduce correctly in certain limits
to the appropriate expressions for the parallel brane case. 
The effective potential constructed from these K\"ahler metrics
is then used to derive soft breaking  under the standard assumptions
of a hidden sector breaking used in supergravity models. The soft
breaking results obtained are in general valid for a wide class
of models since they are given in terms of the general attributes
that would characterize the chiral matter. Remarkably the entire
soft breaking in the sector that involves  the bifundamental
fields can be characterised in terms of indices which are 
the intersection angles measured in units of $\pi$ and are the 
analogue of the twist vectors in the heterotic string case.
These results reduce correctly in certain limits to the soft
breaking  for the  parallel brane case. A weakness of the methods used clearly is the lack 
of a direct derivation of Eq.(\ref{twkae}) from first principles, as well as the complete neglect of twisted 
moduli, that may actually involve important physics. \\

 We have also analyzed 
in this paper the question of gauge coupling unification
and its interconnection with the
constraints that preserve ${\cal N}=1$ supersymmetry. The analysis
indicates that while gauge coupling unification is not generic 
in intersecting brane models, there are regions of the moduli
space where it is possible. Other phenomenological
implications of these models were also explored. Specificallly 
we discussed the constraints on the brane intersection angles 
 from flavor changing
neutral currents and the implications of the CP violating phases
that arise quite naturally in the soft breaking sector. It is
argued that such models have the potential to satisfy the EDM 
constraints and at the same time allow phases which are 
sufficiently large to generate visible effects  in phenomena
at colliders and also  provide sufficient new sources of CP
violation in baryogenesis.  Similarly, the pattern of soft
breaking indicates that such models with R-parity will allow
for dark matter in sufficient amounts to satisfy the current 
astrophysical constraints on cold dark matter. 
 These and other issues deserve further study.


\vspace{.5cm} 
\begin{center}
{\bf Acknowledgements} 
\end{center}

B.~K.~would like to thank Marcus Berg, Dieter L\"ust, Stephan Stieberger, and Jan Troost 
for helpful advice and stimulating discussions. 
The work of B.~K.~was supported by the German Science Foundation (DFG) and in part by
funds provided by the U.S. Department of Energy (D.O.E.) under cooperative research agreement
$\#$DF-FC02-94ER40818. Part of this work was done when P.~N. was visiting 
the Max Planck Institute fur Kernphysik, Heidelberg, Germany, and he thanks the
Max Planck Institute for hospitality and the Alexander von Humboldt Foundation 
for support during this visit. The work of P.~N. was also supported in part by
the U.S. National Science Foundation under the grant NSF-PHY-0139967


\clearpage
\begin{appendix}

\section{ An application of the soft breaking analysis }

We have included this appendix to demonstrate how to compute some of the relevant parameters in practice and 
show how the procedure of converting relative angles into shift vectors works.
The example also shows 
that many problems can arise in the details of any given model. 
We just choose to look at one of the first supersymmetric models constructed in the literature, a 
four generation Standard-like Model given in \cite{Cvetic:2001nr}. The set of brane stacks is defined by the 
following table. 
\begin{center} 
\begin{tabular}{|c|c|c|}
\hline
Sector $[ab]$ & $N_a$ & $({\bf n}_a^{(m)},{\bf m}_a^{(m)})$ \\
\hline
\hline
$A_1$ & 6+2 & (1,1) (1,-2)(1,0)\\
$A_2$ & 2 & (-1,0) (-1,0)(1,0)\\
\hline
$B_1$ & 4 & (1,0) (1,2)(1,-1)\\
$B_2$ & 8 & (1,0) (0,1)(0,-1)\\
\hline
$C_1$ & 2 & (1,2) (1,0)(1,-2)\\
$C_2$ & 8 & (0,1) (1,0)(0,-1)\\
\hline
\end{tabular}
\end{center}
\begin{center} {\bf Table 1}
\end{center}
The constraints of supersymmetry require in this case the relation $\Re(T_1):\Re(T_2):\Re(T_3)=1:2:1$.
Compared to \cite{Cvetic:2001nr} we have  flipped two of the orientations of the stack denoted $A_2$ to achieve 
that all angles $\varphi_a^{(m)}$ add up to $2\pi$ for any stack $a$, according to our conventions. 
All other stacks produce angles 
of the kind $(\alpha , 2\pi-\alpha , 0)$, or permutations thereof, for some values of $\alpha$. 
The spectrum, gauge group and ideas on the physical relevance of this model can be found in the original works. 
The model, as all examples in the literature, does not fix all three K\"ahler parameters $\Re(T_m)$, since in 
Eq.(\ref{susyflux}) the product of fluxes always vanishes for all brane stacks, as one of the winding 
numbers is always zero. Therefore, we can only parametrize the angle variables in terms of the one 
remaining modulus, which we call $\chi=1/\Re(T_1)$. This is somehow against our general philosophy, 
since we neglect the implicit dependence of the shift vectors on the moduli in the soft parameters, but 
cannot be avoided in the absence of more general models, where Eq.(\ref{susyflux}) fixes all moduli. 
Now we apply Eq.(\ref{cos}) to compute the shift vectors, which 
gives Table 2. 
\begin{center} 
\begin{tabular}{|c|c|c|c|}
\hline
Sector $[ab]$ & $\nu_{ab}^{(m)}$ & $\nu_{ab}$ & Fields \\
\hline
\hline
$\nu_{A_1B_1}^{(m)}$ & $(\alpha,2\alpha, \alpha)$ & $4\alpha$ & $Q_L, L$\\
$\nu_{A_1B_2}^{(m)}$ & $(\alpha, \frac{1}{2}+\alpha, \frac{1}{2})$ & $1+2\alpha$ &
$\bar U, \bar D, \bar \nu, \bar E$ \\
$\nu_{A_1C_2}^{(m)}$ & $(\frac{1}{2}-\alpha, \alpha, \frac{1}{2})$ & 1 
& $U,D,\nu, E$\\
$\nu_{B_1C_2}^{(m)}$ & $(\frac{1}{2}, \alpha, \frac{1}{2}-\alpha)$ & 1
& $H_U, H_D$ \\
$\nu_{B_2C_1}^{(m)}$ & $(\alpha',\frac{1}{2}, \frac{1}{2}-\alpha')$ & 1 & $S_U, S_D$ \\
\hline 
\end{tabular}
\end{center}
\begin{center} {\bf Table 2} 
\end{center} 
where $S_U, S_D$  are $SU(2)$ singlets and 
where $\alpha,\alpha'$ are defined by 
\beqn
\alpha &=&\frac{1}{\pi} {\rm arccos} \left( \frac{1}{\sqrt{1+\chi^2}}\right) \ ,\quad 
0\leq \alpha\leq \frac{1}{2} \ , \non
\alpha' &=&\frac{1}{\pi} {\rm arccos} \left( \frac{1}{\sqrt{1+4\chi^2}}\right) \ ,\quad 
0\leq \alpha'\leq \frac{1}{2} \ .
\eeqn
In table 2 we see some non-perturbative sectors ($\nu_{ab}=1$) and 
two ``interpolating'' sectors. 
We notice that this model contains precisely the D9-D5$_1$-D5$_2$ system discussed in \cite{Ibanez:1998rf} and given 
in Eq.(\ref{ibanez2})\footnote{The D5$_3$ brane would need to be defined with $((0,1),(0,1),(-1,0))$.} in the set 
of branes $A_2, B_2, C_2$. Again blindly applying Eq.(\ref{cos}) and get, 

\begin{center} 
\begin{tabular}{|c|c|c|}
\hline
Sector $[ab]$ & $\nu_{ab}^{(m)}$ & $\nu_{ab}$ \\
\hline
\hline
$\nu_{A_2B_2}^{(m)}$ & $(1,1/2,1/2 )$ & 2 \\
$\nu_{A_2C_2}^{(m)}$ & $(1/2,1,1/2)$ & 2 \\
$\nu_{B_2C_2}^{(m)}$ & $(1/2,1/2,0)$ & 1 \\
\hline 
\end{tabular}
\begin{center} {\bf Table 3} 
\end{center} 
\end{center}
exactly reproducing what was given in Eq.(\ref{ibanez2}), via Eq.(\ref{twkae}). 
However, one may note that the procedure is only 
unique once the winding quantum numbers $({\bf n}_a,{\bf m}_a)$ have been specified, 
and switching the orientations differently on the $A_2$ brane would have produced slightly
 different results. 
With the above analysis at  hand we can implement our 
soft breaking formulae for the  4 generation  model. Using the
tables above and the general relations derived in  section 4.1  we get 
\beqn
M^2_{Q_L} &=&
 c^2\left( \frac{1}{3} -\sin^2(\theta) -2\alpha \cos(2\theta)
+\alpha \cos^2(\theta) [1+F_2]  \right) \ , \\ 
M^2_{\bar U, \bar D} &=&
 c^2 \left( -\frac{1}{6} - \alpha \cos(2\theta) 
 + \cos^2(\theta) [\frac{1}{2} + (\alpha-\frac{1}{2}) F_1  + \alpha F_2] \right) \ , \non 
M^2_{U,D} &=&
 c^2\left(-\frac{1}{6} + \frac{1}{2}\cos^2(\theta)  
 + \cos^2(\theta) [-\alpha F_1  +(\alpha-\frac{1}{2})F_2] \right) \ , \non 
M^2_{H_U,H_D} &=&
 c^2\left(-\frac{1}{6} + (\frac{1}{2}-\alpha)\cos^2(\theta)  
 + \cos^2(\theta) [\alpha F_1  +(-\frac{1}{2}+2\alpha)F_2] \right) \ , \non  
%
M^2_{S_U,S_D} &=&
 c^2\left( -\frac{1}{6} + \left( \frac{1}{2}-\alpha' \right)\cos^2(\theta) 
 + \cos^2(\theta) [ (-\frac{1}{2} + 2\alpha' )
  F_1  +  \alpha' F_2] \right) \ , \nonumber
\eeqn
where 
$$
\sum_{i=1}^3 F_i=1,~~F_i = |\Theta_{t_i}|^2+ |\Theta_{u_i}|^2 \ , \quad i=1,2,3
$$
and 
$$
M^2_{Q_L}= M^2_{L}\ , \quad
M^2_{U,D}=M^2_{\nu,E}\ ,\quad 
M^2_{\bar U,\bar D}=M^2_{\bar \nu,\bar E} \ .
$$
Next we focus on the trilinear couplings.
We notice that  the sectors that ``connect'' and have non-vnishing tree level Yukawa couplings are
$[A_1B_1, B_1C_2, C_2A_1]$. 
Thus only the two generations of
 right handed quarks (leptons) in the $A_1C_2$ sector couple
and the other two generations of right handed quarks (leptons) in the sector
$A_1B_2$ do not couple with the left handed quarks (leptons)
and the Higgs. Consequently only two generations of quarks (leptons)
can gain mass by the Higgs phenomenon in this model. In the sector where  the
quark (lepton) mass growth can occur the trilinear couplings are
$$
A^0_{[A_1B_1C_2]} =A^0_{[q_LH_UU]} 
=A^0_{[q_LH_DD]} =
 A^0_{[LH_DE]} =A^0 \ ,
$$
and using the analysis of section 4.1 one has
$$
A^0= -c\frac{e^{-\rho +\frac{D}{2}}}{\sqrt f} \cos (\theta)
(\Theta_{t_2}e^{-i\gamma_{t_2}} +\Theta_{u_2} e^{-i\gamma_{u_2}}) \ .
$$
So for instance, a purely dilaton dominated scenario with $\theta=\pi/2$ would not have any soft 
trilinear couplings in this model.

\end{appendix}



\end{document}